\documentclass{article}

\usepackage[dvips]{graphicx}
\usepackage{cite}

\title{Resonant structure of space-time of early universe}
\author{Sergei~P.~Maydanyuk%
\thanks{\emph{E-mail:} maidan@kinr.kiev.ua} \\
\small\emph{Institute for Nuclear Research, National Academy of Sciences of Ukraine} \\
\small\emph{47, prosp. Nauki, Kiev-28, 03680, Ukraine}}
\date{\small\today}

\begin{document}

\maketitle
%-----------------------------------------------------------------------------------------------------------------------

%-----------------------------------------------------------------------------------------------------------------------
\begin{abstract}
A new fully quantum method describing penetration of packet from internal well outside with its tunneling through the barrier of arbitrary shape used in problems of quantum cosmology, is presented.
The method allows to determine amplitudes of wave function, penetrability $T_{\rm bar}$ and reflection $R_{\rm bar}$ relatively the barrier (accuracy of the method: $|T_{\rm bar}+R_{\rm bar}-1| < 1 \cdot 10^{-15}$), coefficient of penetration (i.~e. probability of the packet to penetrate from the internal well outside with its tunneling), coefficient of oscillations (describing oscillating behavior of the packet inside the internal well).
Using the method, evolution of universe in the closed Friedmann--Robertson--Walker model with quantization in presence of positive cosmological constant, radiation and component of generalize Chaplygin gas is studied.
It is established (for the first time):
(1) oscillating dependence of the penetrability on localization of start of the packet;
(2) presence of resonant values of energy of radiation $E_{\rm rad}$, at which the coefficient of penetration increases strongly.
From analysis of these results it follows:
(1) necessity to introduce initial condition into both non-stationary, and stationary quantum models;
(2) presence of some definite values for the scale factor $a$, where start of expansion of universe is the most probable;
(3) during expansion of universe in the initial stage its radius is changed not continuously, but passes consequently through definite discrete values and tends to continuous spectrum in latter time.
%
% (4) следствием квантования космологической модели является дискретность пространства-времени, проявляющаяся наиболее % сильно в первой стадии расширения вселенной.
\end{abstract}
%-----------------------------------------------------------------------------------------------------------------------

%-----------------------------------------------------------------------------------------------------------------------
\textbf{Keywords:}
physics of the early universe, inflation,
quantum cosmology, Wheeler-De Witt equation, Chaplygin gas, tunneling boundary conditions, penetrability, resonances

\textbf{PACS nubers:}
98.80.Qc, % Quantum cosmology
98.80.–k, % Cosmology
98.80.Bp, % Origin and formation of the Universe, Big Bang theory
98.80.Jk, % Mathematical and relativistic aspects of cosmology
03.65.Xp % Tunneling, traversal time, quantum Zeno dynamics
% 04.60.-m% - Quantum gravity
%-----------------------------------------------------------------------------------------------------------------------

%-----------------------------------------------------------------------------------------------------------------------
% \tableofcontents
% *******************************************************************************************************************

% *******************************************************************************************************************
\section{Introduction
\label{sec.introduction}}

If we analyzed existed variety of quantum models which describe formation of the universe and its subsequent evolution in the first stage, we should come to conclusion that the semiclassical approach for description of tunneling and determination of wave function is the most prevailing today. This approach forms a basis, which props up both models with the Feynman formalism of path integrals in multidimensional space-time, developed by the Cambridge group and other researchers, called the \emph{``Hartle--Hawking method''} (for example, see Ref.~\cite{Hartle.1983.PRD}), and methods based on direct consideration of tunneling in 4-dimensional Euclidian space-time called the \emph{``Vilenkin method''} (for example, see Refs.~\cite{Vilenkin.1982.PLB,Vilenkin.1983.PRD,Vilenkin.1984.PRD,Vilenkin.1986.PRD,Vilenkin.1988.PRD,%
Vilenkin.1994.PRD,Vilenkin.1995,Vilenkin.1997.PRD.Comments,Vilenkin.2003.PRD}). The models in 4-dimensional space-time directed on description of inflation, on study of fluctuations of vacuum with inclusion of massive fields (for example, see~Refs.\cite{Bouhmadi-Lopez.2002.PRD}), multidimensional cosmological models (for example, see~Refs.\cite{Kolb.1990,Carugno.1996.PRD}), variety of string models (for example, see~Refs.\cite{Brustein.2006.PRD,Kobakhidze.2007.EPJC}) have mainly such a semiclassical grounds. To date, this basis is supposed to be sufficiently reliable and give interested quantum characteristics of the universe with good accuracy. Such a point of view so has been taken root and is prevailing, that, in spite of almost 40 years of researches in quantum cosmology (see~the first papers~\cite{DeWitt.1967,Wheeler.1968}), the papers devoted on more tiny and deeper study of possible quantum nature of the universe, its formation and evolution on the first stage, can be found enough rarely (for example, see \cite{Rubakov.2002.PRD,AcacioDeBarros.2007.PRD,Monerat.2007.PRD}, also~\cite{Yurov.2004.TMP,Garcia.2006.IJTP}).

However, one should ask whether the penetrability determined according to the semiclassical theory by a shape of the barrier solely between two turning points, gives exhaustive answer and the best estimations of rates of evolution of universe.

% \vspace{2mm}
(1) If, despite such widespread confidence in the semiclassical approach, we still want to check it, we immediately will miss some of the parameters. For example, it would seem, one can use a test of $T + R = 1$ (where $T$ and $R$ are penetrability and reflection relatively the barrier in the cosmological problem), where usually we have no doubt. However, one should recall that in quantum mechanics, the semiclassical approximation neglects the reflected waves completely (see~\cite{Landau.v3.1989}, eq.~(46.10), p.~205, also p.~221--222) and, therefore, to compare the calculated penetrability $T$ is just not with anything.

% \vspace{2mm}
(2) If we still wanted to determine the reflection coefficient, then we should increase the order of approximation of the semiclassical method (in order to take into account the decreasing component of the wave function on the background of increasing one in the region of tunneling), and just stumble on the next problem --- presence of a non-zero interference between the incident and reflected waves in the radial task. Now the criterion $T + R = 1$ for testing is not satisfied and it needs to take into account the third component $M$ of interference in addition (see~\cite{Maydanyuk.2010.IJMPD}). In particular, at unsuccessfully chosen separation of the exactly known full wave function on the incident and reflected waves (but the semiclassical approaches have no needed apparatus for such analysis), the interference component can increase without limit, and be substantially larger than the penetrability and reflection. After the appearance of such an arbitrariness, the penetrability and reflection can freely exceed the unit and increase without limit. In what is now the general meaning of the penetrability?

% \vspace{2mm}
(3) We shall give only some easy examples from quantum mechanics.
(i) If we consider two-dimensional penetration of the packet through the simplest rectangle barrier (with finite size), we shall see that the penetrability is directly dependent on direction of tunneling of the packet. So, the penetrability is not a single value but the function.
(ii) If to consider one-dimensional tunneling of the packet through the simplest rectangular barrier, one can obtain ``interference picture'' of its amplitude in the transmitted region, which is depended on time and space coordinates and is an exact analytical solution. Of course, the stationary part of such a result coincides exactly with well known stationary solutions \cite{Maydanyuk.2003.PhD-thesis}. From the arguments above the impression could appear that the penetrability defined solely by the shape of the barrier between two turning points is nothing more than prevailing simplified understanding, while for more accurate and deep analysis we need in the strong basis.

% \vspace{2mm}
(4) Advance of the semiclassical approach is in simplicity of formula of the penetrability based on determination of the outgoing wave in the asymptotic region. A \emph{tunneling boundary condition} \cite{Vilenkin.1995,Vilenkin.1994.PRD} seems to be natural and clear, where the wave function should represent an outgoing wave at large scale factor $a$. However, whether is such a wave free? In contrast to problems of quantum atomic and nuclear physics, in cosmology we deals with potentials, which modules only increase with increasing the scale factor $a$ (also their gradients increase, which have sense of force acting on the wave) and, therefore, we have nothing mutual with a free propagation of the wave in the asymptotic region. Now it is unclear to which combination two Airy functions should be combined at turning point, in order to obtain the proper outgoing wave. It turns out that instead of the free wave in the asymptotic (missing in problems of quantum cosmology), we should be able to work with the waves propagating inside strong fields (see~\cite{Maydanyuk.2010.IJMPD}).

% \vspace{2mm}
These problems violate (destroy) the basis of the semiclassical models, and now statements about reliability of the semiclassical models are transformed into the question of `` faith'' in them, though widespread \cite{Maydanyuk.2010.IJMPD}. The semiclassical approach could be compared to \emph{``black box''} in which deeper and more detailed information about the dynamics of the universe is hide. More importantly, in such a black box those missing elements are hidden, without which it is impossible to combine everything together to obtain self-consistent formalism of quantum description of the formation of the universe and its evolution in the first stage. In order to clarify these questions, we have developed a new fully quantum method presented in this paper.

% \vspace{2mm}
This paper is organized so. In Sec.~2 a new fully quantum method for description of the formation of the universe and its evolution in the first stage on the basis of a packet, which penetrates from the internal potential well outside by tunneling through the barrier, is presented. A main advance of this method is determination of characteristics of the packet with high accuracy (without implication of the semiclassical approximations). The formalism of the method has been developed relatively the barrier of arbitrary shape, that makes the method universal. In Sec.~3 the method is applied for solution of the problem of evolution of the packet in FRW-model with radiation and Chaplygin gas. The penetrability and reflection relatively the barrier are calculated. We propose new characteristics, more adequately determining the probability of formation of the universe and its subsequent expansion. An accuracy of the method is demonstrated, achieving to $|T_{\rm bar}+R_{\rm bar}-1| < 1 \cdot 10^{-15}$ inside whole under-barrier range of the energy of radiation ($M = 0$, author has not yet found competitive approaches, achieving such a precision), stability in calculations for the obtained results is shown. A special attention in analysis is devoted to study of the initial conditions. On such a basis, for the first time oscillatory dependence of the penetrability on the localization of start of the packet and resonant levels of the energy of radiation $E_(\rm rad)$ (where the penetration extremely increases) are opened (which are hidden in the semiclassical picture). In finishing, in Sec.~\ref{sec.conclusions} conclusions of the obtained results are formulated.
% *******************************************************************************************************************

% *******************************************************************************************************************
\section{Theoretical approach
\label{sec.model}}

\subsection{A model in the Friedmann--Robertson--Walker metric with radiation and generalized Chaplygin gas
\label{sec.model.1}}

We shall begin from consideration of a closed ($k=1$) FRW model in presence of a positive cosmological constant $\Lambda > 0$, radiation and the component of Chaplygin gas. Let us choose a minisuperspace Lagrangian in the following form
(see Ref.~\cite{Maydanyuk.2010.IJMPD}):
% also Ref.~\cite{Vilenkin.1995}, (11), p.~4):
%
\begin{equation}
\begin{array}{lcl}
  \mathcal{L}\,(a,\dot{a}) =
  \displaystyle\frac{3\,a}{8\pi\,G}\:
  \biggl(-\dot{a}^{2} + k - \displaystyle\frac{8\pi\,G}{3}\; a^{2}\,\rho(a) \biggr), &
  \rho\,(a) =
    \biggl( \rho_{\Lambda} + \displaystyle\frac{\rho_{\rm dust}}{a^{3\,(1+\alpha)}} \biggr)^{1/(1+\alpha)} +
    \displaystyle\frac{\rho_{\rm rad}}{a^{4}(t)}, &
%  \rho\,(a) = \rho_{\Lambda} + \displaystyle\frac{\rho_{\rm rad}}{a^{4}(t)}, &
  \rho_{\Lambda} = \displaystyle\frac{\Lambda}{8\pi\,G},
\end{array}
\label{eq.model.1.1}
\end{equation}
where
$a$ is scale factor,
$\dot{a}$ is derivative of $a$ with respect to time coordinate $t$,
$\rho\,(a)$ is a general expression for the energy density,
$\rho_{\rm rad}(a)$ is component describing the radiation in the initial stage (equation of state for radiation is $p\,(a)=\rho_{\rm rad}(a)/3$, $p$ is pressure),
$\alpha$ is parameter of Chaplygin gas (for details, see Refs.~\cite{Kamenshchik.2001.PLB,Bouhmadi-Lopez.2005.PRD,Bouhmadi-Lopez.2008.IJMPD}, also historical paper \cite{Chaplygin.1904}). The passage to the quantum description of the evolution of the Universe is obtained by the standard procedure of canonical quantization in the Dirac formalism for systems with constraints. In result, we obtain the \emph{Wheeler--De Witt (WDW) equation} (see Ref.~\cite{Vilenkin.1995}, also \cite{Wheeler.1968,DeWitt.1967,Rubakov.2002.PRD}), which after multiplication on factor and passage of the item with radiation $\rho_{\rm rad}$ to right part transforms into the following form (see Ref.~\cite{Maydanyuk.2010.IJMPD}):
\begin{equation}
\begin{array}{cccc}
  \biggl\{ -\:\displaystyle\frac{\partial^{2}}{\partial a^{2}} + V\,(a) \biggr\}\; \varphi(a) =
  E_{\rm rad}\; \varphi(a), &
%   V\,(a) =
%     \biggl( \displaystyle\frac{3}{4\pi\,G} \biggr)^{2}\: k\,a^{2} -
%     \displaystyle\frac{3}{2\pi\,G}\:
%      a^{4}\, \biggl( \rho_{\Lambda} + \displaystyle\frac{\rho_{\rm dust}}{a^{3\,(1+\alpha)}} \biggr)^{1/(1+\alpha)}, &
  E_{\rm rad} = \displaystyle\frac{3\,\rho_{\rm rad}}{2\pi\,G},
\end{array}
\label{eq.model.1.2}
\end{equation}
where
\begin{equation}
\begin{array}{cccc}
  V\,(a) =
    \biggl( \displaystyle\frac{3}{4\pi\,G} \biggr)^{2}\: k\,a^{2} -
    \displaystyle\frac{3}{2\pi\,G}\:
     a^{4}\, \biggl( \rho_{\Lambda} + \displaystyle\frac{\rho_{\rm dust}}{a^{3\,(1+\alpha)}} \biggr)^{1/(1+\alpha)}
\end{array}
\label{eq.model.1.3}
\end{equation}
and $\varphi(a)$ is wave function of Universe. This equation looks similar to one-dimensional stationary Schr\"{o}dinger equation on semiaxis (of variable $a$) at energy $E_{\rm rad}$ with potential $V\,(a)$. It is convenient to use system of units where $8\pi\,G \equiv M_{\rm p}^{-2} = 1$, and
to rewrite $V\,(a)$ in a generalized form as
\begin{equation}
  V(a) =
    A\,a^{2} -
    B\,a^{4}\,\Bigl(\Lambda + \displaystyle\frac{\rho_{\rm dust}}{a^{3\,(1+\alpha)}} \Bigr)^{1/(1+\alpha)},
\label{eq.model.1.4}
\end{equation}
where $A$ and $B$ are constants. In particular, at large $a$ and $A = 36$, $B = 12\,\Lambda$ this potential coincides with \cite{AcacioDeBarros.2007.PRD}.
%-----------------------------------------------------------------------------------------------------------------------

\begin{figure}[h]
\centerline{
\includegraphics[width=93mm]{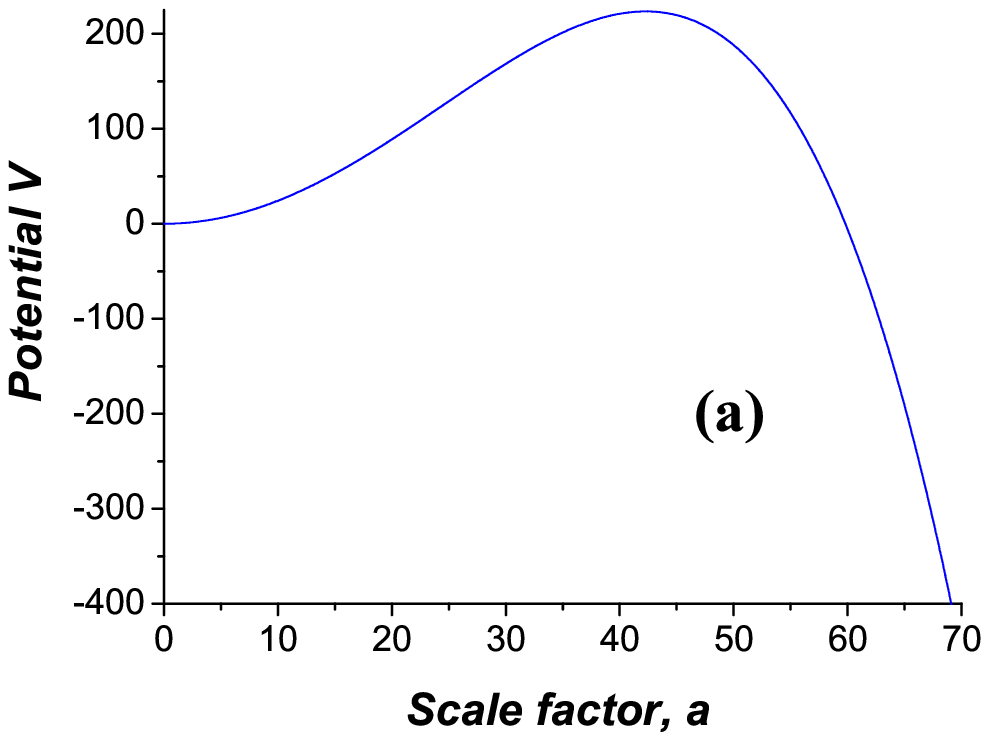}
\hspace{-7mm}\includegraphics[width=93mm]{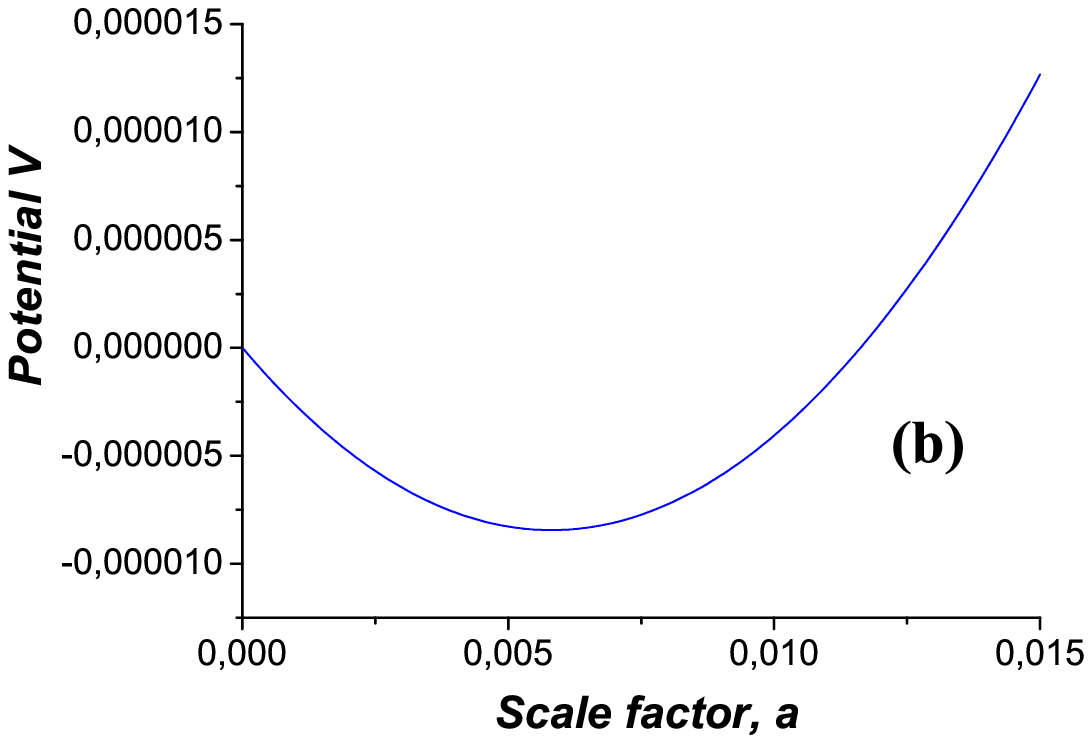}}
\vspace{-1mm}
\caption{\small
Behavior of the potential (\ref{eq.model.2.2}) after change $a_{\rm new} = a_{\rm old}\, \sqrt{12}$ at $\bar{A}=0.001$ and $\bar{B}=0.001$:
(a) shape of the barrier ($V_{\rm max} = 223.52$ at $a=42.322$);
(b) there is a little internal well close to zero ($V_{\rm min} = -8.44$ at $a=0.00581$)
\label{fig.model_Monerat.1}}
\end{figure}

%-----------------------------------------------------------------------------------------------------------------------
In order to estimate ability of the approach developed in this paper below, for comparative analysis let us use results in \cite{Monerat.2007.PRD} where a non-stationary case of the WDW equation
\begin{equation}
  \biggl(
    \displaystyle\frac{1}{12}\,
    \displaystyle\frac{\partial^{2}}{\partial a^{2}} -
    V_{\rm eff}\,(a)
  \biggr)
  \Psi(a, \tau) =
  - i\,\displaystyle\frac {\partial}{\partial \tau}\, \Psi(a, \tau)
\label{eq.model.2.1}
\end{equation}
with the potential for the closed FRW model with the included generalized Chaplygin gas
\begin{equation}
\begin{array}{cc}
\vspace{3mm}
  V_{\rm eff} (a) =
    3\,a^{2} -
    \displaystyle\frac{a^{4}}{\pi}\,
    \sqrt{\bar{A} + \displaystyle\frac{\bar{B}}{a^{6}}}
\end{array}
\label{eq.model.2.2}
\end{equation}
was studied. After change of variable $a_{\rm new} = a_{\rm old}\, \sqrt{12}$ stationary limit of eq.~(\ref{eq.model.2.1}) transforms into our eq.~(\ref{eq.model.1.2}) as the $V_{\rm eff}$ potential is independent on the $\tau$ variable (such a choice keeps correspondence between energy levels that is convenient in comparative analysis).
The potential~(\ref{eq.model.2.2}) after such a transformation is shown in Fig.~\ref{fig.model_Monerat.1}
and we shall study behavior of the wave function concerning it.
% *******************************************************************************************************************

% *******************************************************************************************************************
% \newpage
% \subsection{Penetration of packet through radial barrier
% \label{sec.3}}

\subsection{Tunneling of the packet through the rectangular barrier
\label{sec.3.1}}

Before further solution of our problem, let us consider a general problem of quantum tunneling of a packet through the barrier used in cosmological models. We shall study such a process on the basis of developed formalism of multiple internal reflections (see main ideas, formalism, proof for tunneling, analysis of peculiarities of such approach in
Refs.~\cite{Maydanyuk.2000.UPJ,Maydanyuk.2002.JPS,Maydanyuk.2002.PAST,Maydanyuk.2003.PhD-thesis,%
Maydanyuk.2006.FPL,Maydanyuk.arXiv:0805.4165,Maydanyuk.arXiv:0906.4739},
see also the first papers~\cite{Fermor.1966.AJPIA,McVoy.1967.RMPHA,Anderson.1989.AJPIA}, some development and applications in Refs.~\cite{Esposito.2003.PRE}).
But, at first, we shall start from consideration of main idea of the multiple internal reflections in description of tunneling of the packet on the positive semiaxis of the scale factor $a$ where possible oscillations of such a packet should be included inside internal well and, so, we shall choose the simplest potential $V(a)$ \cite{Maydanyuk.2010.IJMPD}:
$V(a)=-V_{0}$ for $0 < a < R_{1}$ (internal region I),
$V(a)=V_{1}$ for $R_{1} < a < R_{2}$ (region II of the barrier) and
$V(a)=0$ for $a > R_{2}$ (external region III). % (see Fig.~\ref{fig.app.4}).
%
% \begin{figure}[htbp]
% \centerline{\includegraphics[width=50mm]{Figures/Mir_fig_3_2.eps}}
% \caption{\small
% Propagation of wave packet from the internal region outside with its possible tunneling through the barrier
% \label{fig.app.4}}
% \end{figure}
%
For simplicity, let us analyze a case when total energy of system $E$ is higher then the barrier height $V_{1}$: $E>V_{1}$.

In the first step we consider start of the packet in the region I which is assumed to propagate to the right and
is incident on the first boundary of the barrier at $R_{1}$:
\begin{equation}
\begin{array}{lcll}
  \psi_{\rm inc}^{(1)}(a, t) & = &
    \int\limits_{E_{\rm min}}^{+\infty} g(E - \bar{E})\: e^{ik_{1}a -iEt/\hbar}\; dE
    & \mbox{at } 0<a<R_{1},
\end{array}
\label{eq.3.1.1}
\end{equation}
where $k_{1} = \sqrt{E+V_{0}}$, $E$ is the energy.
The weight amplitude $g(E - \bar{E})$ can be used in standard form of gaussian and satisfies to normalization $\int |g(E - \bar{E})|^{2}\: dE = 1$, value $\bar{E}$ is an average energy.
This packet transforms into two new packets: the first packet transmitted through this boundary and propagating further in the region II,
and the second one reflected from the boundary and propagating back in the region I:
\begin{equation}
\begin{array}{lcll}
  \psi^{(1)}_{\rm tr}(a, t) & = &
    \int\limits_{E_{\rm min}}^{+\infty} g(E - \bar{E})\, \alpha^{(1)}\, e^{ia_{2}a -iEt/\hbar}\; dE &
    \mbox{at } R_{1}<a<R_{2}, \\
  \psi^{(1)}_{\rm ref}(a, t) & = &
    \int\limits_{E_{\rm min}}^{+\infty} g(E - \bar{E})\, A_{R}^{(1)}\, e^{-ik_{1}a -iEt/\hbar}\; dE
        & \mbox{at } 0<a<R_{1},
\end{array}
\label{eq.3.1.2}
\end{equation}
where $k_{2} = \sqrt{E-V_{1}}$.
We find new unknown coefficients $\alpha^{(1)}$ and $A_{R}^{(1)}$, using requirements of continuity of the total wave function $\psi(a,t)$ (which is summation of all packets)
and its derivative at $R_{1}$:
\begin{equation}
\begin{array}{ll}
  \alpha^{(1)} = \displaystyle\frac{2\,k_{1}}{k_{1}+k_{2}}\, e^{i\,(k_{1}-k_{2})\,R_{1}}, &
  A_{R}^{(1)}  = \displaystyle\frac{k_{1}-k_{2}}{k_{1}+k_{2}}\, e^{2i\,k_{1}\,R_{1}}.
\end{array}
\label{eq.3.1.3}
\end{equation}

In the second step we consider further propagation of the packet $\psi^{(1)}_{\rm tr}(a, t)$, which is incident on the second boundary at $R_{2}$. It transforms into two new packets: the first packet transmitted through this boundary and propagating in the region III, and the second one reflected from this boundary and propagating back in the region II.
We define these packets in the form
\begin{equation}
\begin{array}{lcl}
  \psi^{(2)}(a, t) & = &
    \int\limits_{E_{\rm min}}^{+\infty} g(E - \bar{E})\: \varphi^{(2)}(a)\: e^{-iEt/\hbar}\; dE,
\end{array}
\label{eq.3.1.4}
\end{equation}
where as the stationary parts we use:
\begin{equation}
\begin{array}{lcll}
\varphi_{\rm inc}^{(2)}(a) & = & \alpha^{(1)} e^{ik_{2}a}
        & \mbox{at } R_{1}<a<R_{2}, \\
\varphi_{\rm tr}^{(2)}(a) & = & A_{T}^{(1)}e^{ika}
        & \mbox{at } a>R_{2}, \\
\varphi_{\rm ref}^{(2)}(a) & = & \beta^{(1)} e^{-ik_{2}a}
        & \mbox{at } R_{1}<a<R_{2},
\end{array}
\label{eq.3.1.5}
\end{equation}
where $k = \sqrt{E}$. Imposing condition of continuity on the total wave function and its derivative at $R_{2}$, we obtain two new equations,
from which we find new unknowns coefficients $A_{T}^{(1)}$ and $\beta^{(1)}$:
\begin{equation}
\begin{array}{ll}
  A_{T}^{(1)} = T_{2}^{+} \cdot \alpha^{(1)}, &
  T_{2}^{+} = \displaystyle\frac{2\,k_{2}}{k_{2}+k}\, e^{i\,(k_{2}-k)\,R_{2}}, \\
  \beta^{(1)} = R_{2}^{+} \cdot \alpha^{(1)}, &
  R_{2}^{+} = \displaystyle\frac{k_{2}-k}{k_{2}+k}\, e^{2i\,k_{2}\,R_{2}}.
\end{array}
\label{eq.3.1.6}
\end{equation}
We have introduced two new coefficients $T_{2}^{+}$, $R_{2}^{+}$, which logically connect the transmitted and reflected amplitudes $A_{T}^{(1)}$ and $\beta^{(1)}$ with the incident amplitude $\alpha^{(1)}$ in this step. Here we shall use bottom index for denotation of number of the considered boundary, upper (top) sign ``$+$'' or ``$-$'' for positive (to the right) or negative (to the left) direction of the incident wave, correspondingly.
So, we can write $T_{1}^{+} = \alpha^{(1)}$ and $R_{1}^{+} = A_{R}^{(1)}$ also.

In the third step we consider further propagation of the reflected packet $\psi^{(2)}_{\rm ref}$ in the region II. Incidenting on the first boundary, it transforms into new packet $\psi^{(3)}_{\rm tr}$, transmitted through this boundary and propagating in the region I, and into new packet $\psi^{(3)}_{\rm ref}$, reflected from boundary and propagating back in the region II. We define the new packets by eq.~(\ref{eq.3.1.4}) (with upper index 3),
where as the stationary parts we use:
\begin{equation}
\begin{array}{lcll}
  \varphi_{\rm inc}^{(3)}(a) & = & \varphi_{\rm ref}^{(2)}(a)
          & \mbox{at } R_{1}<a<R_{2}, \\
  \varphi_{\rm tr}^{(3)}(a) & = & A_{R}^{(2)}e^{-ik_{1}a}
          & \mbox{at } 0<a<R_{1}, \\
  \varphi_{\rm ref}^{(3)}(a) & = & \alpha^{(2)} e^{ik_{2}a}
          & \mbox{at } R_{1}<a<R_{2}.
\end{array}
\label{eq.3.1.7}
\end{equation}
From continuity conditions for the total wave function and its derivative at $R_{1}$ we find the unknowns coefficients
$A_{R}^{(2)}$ and $\alpha^{(2)}$:
\begin{equation}
\begin{array}{ll}
  A_{T}^{(2)}  = T_{1}^{-} \cdot \beta^{(1)}, &
  T_{1}^{-} = \displaystyle\frac{2\,k_{2}}{k_{1}+k_{2}}\, e^{i\,(k_{1}-k_{2})\,R_{1}}, \\
  \alpha^{(2)} = R_{1}^{-} \cdot \beta^{(1)}, &
  R_{1}^{-} = \displaystyle\frac{k_{2}-k_{1}}{k_{1}+k_{2}}\, e^{-2i\,k_{2}\,R_{1}}.
\end{array}
\label{eq.3.1.8}
\end{equation}

In the forth step we need to consider further propagation of the reflected packet $\psi_{\rm ref}^{(1)}$ in the region I in the 1-st step. It is incident on the first boundary at $a=0$ transforming into new packet propagated to the right. At such a point we can include different considerations of origin of possible sources at $a=0$, possible full propagation (like in spherically symmetric problems of quantum decay in nuclear physics which is 3-dimensional and we have no additional boundaries at $a=0$) or, in contrary, possible full reflection used in different quantum approaches (like introduction of an infinite potential wall at $a=0$ in Ref.~\cite{AcacioDeBarros.2007.PRD}).
In order to produce ability to work with different such considerations, we write:
\begin{equation}
\begin{array}{lll}
  \varphi_{\rm inc}^{(4)}(a) = \varphi_{\rm ref}^{(1)}(a), &
  \varphi_{\rm tr}^{(4)}(a, k_{1}) = R_{0}^{-} \cdot \varphi_{\rm inc}^{(4)}(a,-k_{1}) =
    A_{\rm ref}^{(4)}\,e^{ik_{1}a} &
  \mbox{at } 0<a<R_{1},
\end{array}
\label{eq.3.1.9}
\end{equation}
where
\begin{equation}
  A_{\rm ref}^{(4)} = R_{0}^{-} \cdot A_{\rm inc}^{(1)}.
\label{eq.3.1.10}
\end{equation}
Supposing full propagation through this boundary (without any possible reflections), we obtain $R_{0}^{-} = -1$.
If we liked to include the infinite potential wall at $a=0$, than we should have $R_{0}^{-} = -1$ also.

Analyzing further reflections and transmission of the packets concerning the boundaries by such a way, we have concluded that any following step is similar to one from 4 considered above. From analysis of these steps we have found recurrent relations for calculation of unknown amplitudes $A_{\rm inc}^{(n)}$, $A_{R}^{(n)}$, $A_{T}^{(n)}$ $\alpha^{(n)}$ and $\beta^{(n)}$ for arbitrary step $n$, and we have calculated summations of these amplitudes. However, these series could be calculated easier, if to apply coefficients $T_{i}^{\pm}$ and $R_{i}^{\pm}$. Analyzing all possible ``paths''
of the propagations of all possible packets inside the barrier and internal well, we obtain:
\begin{equation}
\begin{array}{lcl}
  \sum\limits_{n=1}^{+\infty} A_{\rm inc}^{(n)} & = &
    1 + \tilde{R}_{1}^{+}\,R_{0}^{-} + \tilde{R}_{1}^{+}\,R_{0}^{-} \cdot \tilde{R}_{1}^{+}\,R_{0}^{-} + ... =
    1 + \sum\limits_{m=1}^{+\infty} \bigl(\tilde{R}_{1}^{+}\,R_{0}^{-}\bigr)^{m} =
    \displaystyle\frac{1}{1 - \tilde{R}_{1}^{+}\,R_{0}^{-}}, \\

  \sum\limits_{n=1}^{+\infty} A_{T}^{(n)} & = &
  \Bigl( \sum\limits_{n=1}^{+\infty} A_{\rm inc}^{(n)} \Bigr) \cdot
  \Bigl\{ T_{1}^{+}\,T_{2}^{+} + T_{1}^{+}\cdot R_{2}^{+}\,R_{1}^{-}\cdot T_{2}^{+} + ... \Bigr\} =
%        T_{1}^{+}\cdot R_{2}^{+}\,R_{1}^{-}\cdot R_{2}^{+}\,R_{1}^{-} \cdot T_{2}^{+} + ...\Bigr\} = % \\
% \vspace{2mm}
%   & = &
  \Bigl( \sum\limits_{n=1}^{+\infty} A_{\rm inc}^{(n)} \Bigr) \cdot \tilde{T}_{2}^{+}, \\

  \sum\limits_{n=1}^{+\infty} A_{R}^{(n)} & = &
    \tilde{R}_{1}^{+} + \tilde{R}_{1}^{+} \cdot R_{0}^{-}\,\tilde{R}_{1}^{+} +
    \tilde{R}_{1}^{+} \cdot R_{0}^{-}\,\tilde{R}_{1}^{+} \cdot R_{0}^{-}\,\tilde{R}_{1}^{+} + ... = \\
\vspace{2mm}
  & = &
    \tilde{R}_{1}^{+} \cdot
      \Bigl( 1 + \sum\limits_{m=1}^{+\infty} \bigl(R_{0}^{-}\,\tilde{R}_{1}^{+}\bigr)^{m} \Bigr) =
    \displaystyle\frac{\tilde{R}_{1}^{+}} {1 - R_{0}^{-}\,\tilde{R}_{1}^{+}} =
  \Bigl( \sum\limits_{n=1}^{+\infty} A_{\rm inc}^{(n)} \Bigr) \cdot \tilde{R}_{1}^{+},
\end{array}
\label{eq.3.1.11}
\end{equation}
where
\begin{equation}
\begin{array}{lcl}
  \tilde{R}_{1}^{+} & = &
%    R_{1}^{+} + T_{1}^{+}\,R_{2}^{+}\,T_{1}^{-} +
%     T_{1}^{+}\,R_{2}^{+}\cdot R_{1}^{-}\, R_{2}^{+} \cdot T_{1}^{-} + ...
%     T_{1}^{+}\,R_{2}^{+}\cdot \bigl(R_{1}^{-}\, R_{2}^{+}\bigr)^{m} \cdot T_{1}^{-} + ... = \\
%   & = &
    R_{1}^{+} + T_{1}^{+}\,R_{2}^{+}\,T_{1}^{-} \cdot
      \Bigl( 1 + \sum\limits_{m=1}^{+\infty} \bigl(R_{2}^{+}\,R_{1}^{-}\bigr)^{m} \Bigr) =
    R_{1}^{+} + \displaystyle\frac{T_{1}^{+}\,R_{2}^{+}\,T_{1}^{-}} {1 - R_{2}^{+}\,R_{1}^{-}}, \\

  \tilde{T}_{2}^{+} & = &
    T_{1}^{+}\, T_{2}^{+} \cdot
    \Bigl( 1 + \sum\limits_{m=1}^{+\infty} \bigl(R_{2}^{+}\,R_{1}^{-}\bigr)^{m} \Bigr) =
    \displaystyle\frac{T_{1}^{+}\, T_{2}^{+}}{1 - R_{2}^{+}\,\tilde{R}_{1}^{-}}.
\end{array}
\label{eq.3.1.12}
\end{equation}

% \begin{equation}
% \begin{array}{lcl}
%   \tilde{T}_{2}^{+} & = &
%   \tilde{T}_{1}^{+} \cdot
%   \Bigl\{ T_{2}^{+} + R_{2}^{+}\,\tilde{R}_{1}^{-}\cdot T_{2}^{+} +
%         R_{2}^{+}\,\tilde{R}_{1}^{-} \cdot R_{2}^{+}\,\tilde{R}_{1}^{-} \cdot T_{2}^{+} + ... +
%         \bigl(R_{2}^{+}\,\tilde{R}_{1}^{-} \bigr)^{m} \cdot T_{2}^{+} + ...\Bigr\} = \\
% \vspace{2mm}
%   & = &
%   \tilde{T}_{1}^{+}\, T_{2}^{+} \cdot
%     \Bigl( 1 + \sum\limits_{m=1}^{+\infty} \bigl(R_{2}^{+}\,\tilde{R}_{1}^{-}\bigr)^{m} \Bigr) =
%     \displaystyle\frac{\tilde{T}_{1}^{+}\, T_{2}^{+}}{1 - R_{2}^{+}\,\tilde{R}_{1}^{-}}, \\
%
%   \tilde{T}_{1}^{+} & = &
%     T_{1}^{+} + R_{1}^{+}\,R_{0}^{-}\cdot T_{1}^{+} +
%     R_{1}^{+}\,R_{0}^{-} \cdot R_{1}^{+}\,R_{0}^{-} \cdot T_{1}^{+} + ... +
%     \bigl(R_{1}^{+}\,R_{0}^{-} \bigr)^{m} \cdot T_{1}^{+} + ... = \\
% \vspace{2mm}
%   & = &
%   T_{1}^{+} \cdot
%     \Bigl( 1 + \sum\limits_{m=1}^{+\infty} \bigl(R_{1}^{+}\,R_{0}^{-}\bigr)^{m} \Bigr) =
%     \displaystyle\frac{T_{1}^{+}}{1 - R_{1}^{+}\,R_{0}^{-}}, \\
%
%   \tilde{R}_{1}^{-} & = &
%     R_{1}^{-} + T_{1}^{-}\,R_{0}^{-}\, T_{1}^{+} +
%     T_{1}^{-}\,R_{0}^{-} \cdot R_{1}^{+}\,R_{0}^{-}\cdot T_{1}^{+} + ... +
%     T_{1}^{-}\,R_{0}^{-} \cdot \bigl(R_{1}^{+}\,R_{0}^{-}\bigr)^{m} \cdot T_{1}^{+} + ... = \\
%   & = &
%     R_{1}^{-} + T_{1}^{-}\,R_{0}^{-}\, T_{1}^{+} \cdot
%     \Bigl( 1 + \sum\limits_{m=1}^{+\infty} \bigl(R_{0}^{-}\,R_{1}^{+}\bigr)^{m} \Bigr) =
%     R_{1}^{-} + \displaystyle\frac{T_{1}^{-}\,R_{0}^{-}\, T_{1}^{+}}{1 - R_{0}^{-}\,R_{1}^{+}},
% \end{array}
% \label{eq.3.1.12}
% \end{equation}

The resultant expressions for the incident, transmitted and reflected packets concerning the barrier are written in form of eq.~(\ref{eq.3.1.4}),
where the following stationary wave functions should be used:
\begin{equation}
\begin{array}{lcll}
\varphi_{\rm inc}(a) & = & e^{ik_{1}a},
                        & \mbox{for } 0<a<R_{1}, \\
\varphi_{\rm tr}(a)  & = & \sum\limits_{n=0}^{+\infty} A_{T}^{n} e^{ika},
                        & \mbox{for } a>R_{2}, \\
\varphi_{\rm ref}(a) & = & \sum\limits_{n=0}^{+\infty} A_{R}^{n} e^{-ik_{1}a},
                        & \mbox{for } 0<a<R_{1}.
\end{array}
\label{eq.3.1.13}
\end{equation}
At finishing, we determine the full amplitudes
\begin{equation}
\begin{array}{cccc}
  A_{T} = \sum\limits_{n=1}^{+\infty} A_{T}^{(n)}, &
  A_{R} = \sum\limits_{n=1}^{+\infty} A_{R}^{(n)}, &
  \alpha = \sum\limits_{n=1}^{+\infty} \alpha^{(n)} =
  \displaystyle\frac{\tilde{T}_{2}^{+}} {T_{2}^{+}}, &
%  \alpha_{j} = \displaystyle\frac{A_{T}}{T_{2}^{+}}, &
  \beta = \sum\limits_{n=1}^{+\infty} \beta^{(n)} = \alpha \cdot R_{2}^{+}
\end{array}
\label{eq.3.1.14}
\end{equation}
and coefficients $T$ and $R$ describing penetration of the packet from the internal region outside
and its reflection from the barrier
\begin{equation}
\begin{array}{ll}
\vspace{1mm}
  T_{MIR} \equiv \displaystyle\frac{k}{k_{1}}\; \bigl|A_{T}\bigr|^{2} =
    \bigl|A_{\rm inc}\bigr|^{2} \cdot T_{\rm bar}, &
  T_{\rm bar} = \displaystyle\frac{k}{k_{1}}\; \bigl|\tilde{T}_{1}^{+} \bigr|^{2}, \\
  R_{MIR} \equiv \bigl|A_{R}\bigr|^{2} = \bigl|A_{\rm inc}\bigr|^{2} \cdot R_{\rm bar}, &
  R_{\rm bar} = \bigl|\tilde{R}_{1}^{+} \bigr|^{2},
\end{array}
\label{eq.3.1.15}
\end{equation}
where $T_{\rm bar}$ and $R_{\rm bar}$ are coefficients of penetrability and reflection of the barrier (in standard definition), and $\bigl|A_{\rm inc}\bigr|^{2}$ is coefficient determining oscillations of the packet inside the internal region (this is fully quantum analog of the normalization factor $F$ introduced in Ref.~\cite{Gurvitz.1987.PRL} for semiclassical description of nuclear decay).

Series $\sum A_{\rm inc}^{(n)}$, $\sum A_{T}^{(n)}$, $\sum A_{R}^{(n)}$, $\sum \alpha^{(n)}$ and $\sum \beta^{(n)}$ obtained using the approach of the multiple internal reflections, \underline{exactly} coincide with the corresponding coefficients $A_{\rm inc}$, $A_{T}$, $A_{R}$, $\alpha$ and $\beta$ calculated by standard stationary method (where the continuity conditions of the stationary total wave function and its derivative are used at each boundaries, and the wave function is not equal to zero at $a=0$).
We test property:
\begin{equation}
\begin{array}{ccc}
  \displaystyle\frac{k}{k_{1}}\; |A_{T}|^{2} + |A_{R}|^{2} = 1 & \mbox{ or }&
  T_{MIR} + R_{MIR} = 1,
\end{array}
\label{eq.3.1.16}
\end{equation}
which is fulfilled and confirms that the method MIR gives us proper solution for the wave function.
If energy is less then the height of the barrier, then for description of penetration of the wave through such a barrier with its tunneling it needs to use
the following change~\cite{Maydanyuk.2002.JPS,Maydanyuk.2003.PhD-thesis,Maydanyuk.2006.FPL}:
\begin{equation}
\begin{array}{cc}
  k_{2} \to i\,\xi, &
  \xi = \sqrt{E-V_{1}}.
\end{array}
\label{eq.3.1.17}
\end{equation}
Using it, all found above solutions are applied for the problem with tunneling through the barrier.
%-----------------------------------------------------------------------------------------------------------------------

%-----------------------------------------------------------------------------------------------------------------------
\subsection{Tunneling of the packet through barrier composed from arbitrary number of rectangular steps
\label{sec.3.2}}

Now let us come to another essentially more difficult problem of the packet penetrating through the radial barrier of arbitrary shape in the cosmological problem. In order to apply the idea of multiple internal refections for study the packet tunneling through the real barrier, we have to generalize formalism of the multiple internal reflections presented above.
We shall assume that the total potential has successfully been approximated by finite number $N$ of rectangular steps:
\begin{equation}
  V(a) = \left\{
  \begin{array}{cll}
    V_{1},   & \mbox{at } a_{\rm min} < a \leq a_{1}      & \mbox{(region 1)}, \\
    V_{2},   & \mbox{at } a_{1} < a \leq a_{2}         & \mbox{(region 2)}, \\
%    V_{3},   & \mbox{at } a_{2} < a \leq a_{3}        & \mbox{(region 3)}, \\
    \ldots   & \ldots & \ldots \\
%    V_{N-1}, & \mbox{at } a_{N-2} < a \leq a_{N-1}     & \mbox{(region $N-1$)}, \\
    V_{N},   & \mbox{at } a_{N-1} < a \leq a_{\rm max} & \mbox{(region $N$)},
  \end{array} \right.
\label{eq.3.2.1}
\end{equation}
where $V_{i}$ are constants ($i = 1 \ldots N$).
Now let us assume that the packet starts to propagate outside inside the region with some arbitrary number $M$ (for simplicity, we denote its left boundary $a_{M-1}$ as $a_{\rm start}$) from the left of the barrier. We shall be interesting in solutions for above barrier energies while the solution for tunneling could be obtained after by change $i\,\xi_{i} \to k_{i}$. A general solution of the wave function (up to
its normalization) has the following form:
\begin{equation}
\varphi\,(a) = \left\{
\begin{array}{lll}
   \alpha_{1}\, e^{ik_{1}a} + \beta_{1}\, e^{-ik_{1}a}, &
     \mbox{at } a_{\rm min} \leq a \leq a_{1} & \mbox{(region 1)}, \\
   \ldots & \ldots & \ldots \\
   \alpha_{M-1}\, e^{ik_{M-1}a} + \beta_{M-1}\, e^{-ik_{M-1}a}, &
     \mbox{at } a_{M-2} \leq a \leq a_{\rm M-1} & \mbox{(region $M-1$)}, \\

   e^{ik_{M}a} + A_{R}\,e^{-ik_{M}a}, & \mbox{at } a_{\rm M-1} < a \leq a_{M} & \mbox{(region $M$)}, \\
   \alpha_{M+1}\, e^{ik_{M+1}a} + \beta_{M+1}\, e^{-ik_{M+1}a}, &
     \mbox{at } a_{M} \leq a \leq a_{M+1} & \mbox{(region $M+1$)}, \\
%    \alpha_{3}\, e^{ik_{3}a} + \beta_{3}\, e^{-ik_{3}a}, & \mbox{at } a_{2} \leq a \leq a_{3} & \mbox{(region 3)}, \\
   \ldots & \ldots & \ldots \\
   \alpha_{n-1}\, e^{ik_{N-1}a} + \beta_{N-1}\, e^{-ik_{N-1}a}, &
     \mbox{at } a_{N-2} \leq a \leq a_{N-1} & \mbox{(region $N-1$)}, \\
   A_{T}\,e^{ik_{N}a}, & \mbox{at } a_{N-1} \leq a \leq a_{\rm max} & \mbox{(region $N$)},
\end{array} \right.
\label{eq.3.2.2}
\end{equation}
where $\alpha_{j}$ and $\beta_{j}$ are unknown amplitudes, $A_{T}$ and $A_{R}$ are unknown amplitudes of transmission and reflection, $k_{i} = \frac{1}{\hbar}\sqrt{2m(E-V_{i})}$ are complex wave numbers. We have fixed the normalization so that modulus of the starting wave $e^{ik_{M}a}$ equals to one. We shall be looking for solution for such a problem by the approach of the multiple internal reflections.

Let us begin from consideration of start of the packet in the region with number $M$, which propagates to the right. At first, we shall study its propagation inside the right part of the potential with barrier, starting from this region. According to the method of the multiple internal reflections, scattering of the packet on the barrier is considered consequently by steps of its propagation relatively to each boundary of the barrier (the most clearly idea of such approach can be understood in the problem of tunneling through the simplest rectangular barrier, see \cite{Maydanyuk.2002.JPS,Maydanyuk.2003.PhD-thesis,Maydanyuk.2006.FPL} where one can find proof of this fully quantum exactly solvable method, one can analyze its properties).
Each step in such consideration of propagation of the packet will be similar to one from the first $2N-1$ steps, independent between themselves. From analysis of these steps recurrent relations are found for calculation of unknown amplitudes $A^{(n)}$, $S^{(n)}$, $\alpha^{(n)}$ and $\beta^{(n)}$ for arbitrary step $n$, summation of these amplitudes are calculated.
We shall be looking for the unknown amplitudes, requiring wave function and its derivative to be continuous at each boundary. We shall consider the coefficients $T_{1}^{\pm}$, $T_{2}^{\pm}$ \ldots and $R_{1}^{\pm}$, $R_{2}^{\pm}$ \ldots as additional factors to amplitudes $e^{\pm i\,k\,a}$. Here, bottom index denotes number of the region, upper (top) signs ``$+$'' and ``$-$'' denote directions of the wave to the right or to the left, correspondingly. At the first, we calculate $T_{1}^{\pm}$, $T_{2}^{\pm}$ \ldots $T_{N-1}^{\pm}$ and $R_{1}^{\pm}$,
$R_{2}^{\pm}$ \ldots $R_{N-1}^{\pm}$:
\begin{equation}
\begin{array}{ll}
\vspace{2mm}
   T_{j}^{+} = \displaystyle\frac{2k_{j}}{k_{j}+k_{j+1}} \,e^{i(k_{j}-k_{j+1}) a_{j}}, &
   T_{j}^{-} = \displaystyle\frac{2k_{j+1}}{k_{j}+k_{j+1}} \,e^{i(k_{j}-k_{j+1}) a_{j}}, \\
   R_{j}^{+} = \displaystyle\frac{k_{j}-k_{j+1}}{k_{j}+k_{j+1}} \,e^{2ik_{j}a_{j}}, &
   R_{j}^{-} = \displaystyle\frac{k_{j+1}-k_{j}}{k_{j}+k_{j+1}} \,e^{-2ik_{j+1}a_{j}}.
\end{array}
\label{eq.3.2.3}
\end{equation}
Analyzing all possible ``paths'' of the propagations of all possible packets inside the barrier and internal well,
we obtain:
\begin{equation}
\begin{array}{lcl}
  \sum\limits_{n=1}^{+\infty} A_{\rm inc}^{(n)} & = &
    1 + \tilde{R}_{M}^{+}\,\tilde{R}_{M-1}^{-} +
    \tilde{R}_{M}^{+}\,\tilde{R}_{M-1}^{-} \cdot \tilde{R}_{M}^{+}\,\tilde{R}_{M-1}^{-} + ... =
    1 + \sum\limits_{m=1}^{+\infty} \bigl(\tilde{R}_{M}^{+}\,\tilde{R}_{M-1}^{-}\bigr)^{m} =
    \displaystyle\frac{1}{1 - \tilde{R}_{M}^{+}\,\tilde{R}_{M-1}^{-}}, \\

  \sum\limits_{n=1}^{+\infty} A_{T}^{(n)} & = &
  \Bigl( \sum\limits_{n=1}^{+\infty} A_{\rm inc}^{(n)} \Bigr) \cdot
  \Bigl\{ \tilde{T}_{N-2}^{+}\,T_{N-1}^{+} +
         \tilde{T}_{N-2}^{+}\cdot R_{N-1}^{+}\,\tilde{R}_{N-2}^{-}\cdot T_{N-1}^{+} + ... \Bigr\} =
  \Bigl( \sum\limits_{n=1}^{+\infty} A_{\rm inc}^{(n)} \Bigr) \cdot \tilde{T}_{N-1}^{+}, \\

  \sum\limits_{n=1}^{+\infty} A_{R}^{(n)} & = &
    \tilde{R}_{M}^{+} + \tilde{R}_{M}^{+} \cdot \tilde{R}_{M-1}^{-}\,\tilde{R}_{M}^{+} +
    \tilde{R}_{M}^{+} \cdot \tilde{R}_{M-1}^{-}\,\tilde{R}_{M}^{+} \cdot \tilde{R}_{M-1}^{-}\,\tilde{R}_{M}^{+} + ... =\\
\vspace{2mm}
  & = &
    \tilde{R}_{M}^{+} \cdot
      \Bigl( 1 + \sum\limits_{m=1}^{+\infty} \bigl(\tilde{R}_{M-1}^{-}\,\tilde{R}_{M}^{+}\bigr)^{m} \Bigr) =
    \displaystyle\frac{\tilde{R}_{M}^{+}} {1 - \tilde{R}_{M-1}^{-}\,\tilde{R}_{M}^{+}} =
  \Bigl( \sum\limits_{n=1}^{+\infty} A_{\rm inc}^{(n)} \Bigr) \cdot \tilde{R}_{M}^{+},
\end{array}
\label{eq.3.2.4}
\end{equation}
where
\begin{equation}
\begin{array}{l}
   \vspace{1mm}
   \tilde{R}_{j-1}^{+} =
     R_{j-1}^{+} + T_{j-1}^{+} \tilde{R}_{j}^{+} T_{j-1}^{-}
     \Bigl(1 + \sum\limits_{m=1}^{+\infty} (\tilde{R}_{j}^{+}R_{j-1}^{-})^{m} \Bigr) =
     R_{j-1}^{+} +
     \displaystyle\frac{T_{j-1}^{+} \tilde{R}_{j}^{+} T_{j-1}^{-}} {1 - \tilde{R}_{j}^{+} R_{j-1}^{-}}, \\

   \vspace{1mm}
   \tilde{R}_{j+1}^{-} =
     R_{j+1}^{-} + T_{j+1}^{-} \tilde{R}_{j}^{-} T_{j+1}^{+}
     \Bigl(1 + \sum\limits_{m=1}^{+\infty} (R_{j+1}^{+} \tilde{R}_{j}^{-})^{m} \Bigr) =
     R_{j+1}^{-} +
     \displaystyle\frac{T_{j+1}^{-} \tilde{R}_{j}^{-} T_{j+1}^{+}} {1 - R_{j+1}^{+} \tilde{R}_{j}^{-}}, \\

   \tilde{T}_{j+1}^{+} =
     \tilde{T}_{j}^{+} T_{j+1}^{+}
     \Bigl(1 + \sum\limits_{m=1}^{+\infty} (R_{j+1}^{+} \tilde{R}_{j}^{-})^{m} \Bigr) =
     \displaystyle\frac{\tilde{T}_{j}^{+} T_{j+1}^{+}} {1 - R_{j+1}^{+} \tilde{R}_{j}^{-}},
\end{array}
\label{eq.3.2.5}
\end{equation}
and selecting as starting the following values:
\begin{equation}
\begin{array}{ccc}
  \tilde{R}_{N-1}^{+} = R_{N-1}^{+}, &
  \tilde{R}_{M}^{-} = R_{M}^{-}, &
  \tilde{T}_{M}^{+} = T_{M}^{+},
\end{array}
\label{eq.3.2.6}
\end{equation}
we calculate successively coefficients $\tilde{R}_{N-2}^{+}$ \ldots $\tilde{R}_{M}^{+}$, $\tilde{R}_{M+1}^{-}$ \ldots
$\tilde{R}_{N-1}^{-}$ and $\tilde{T}_{M+1}^{+}$ \ldots $\tilde{T}_{N-1}^{+}$.

At second, we shall consider further propagation of all packets which propagate in the region with number $M$ to the left. Such packets are formed in result of all possible reflections from the right part of potential, starting from the boundary $a_{M}$. In the previous section for description of their reflection from the left boundary $R_{0}$ to the right we used coefficient $R_{0}^{-}$. Now if we liked to pass from simple boundary $a_{M-1}$ to the left part of the potential well starting from this point up to $a_{\rm min}$, we should generalize the coefficient $R_{M-1}^{-}$ on $\tilde{R}_{M-1}^{-}$. It turns out that the middle recurrent formula (\ref{eq.3.2.5}) is absolutely applicable in such a case also, where for definition of $T_{i}^{\pm}$ and $R_{i}^{\pm}$ we should use eqs.~(\ref{eq.3.2.3}) again.
At finishing, we determine coefficients $\alpha_{j}$ and $\beta_{j}$:
\begin{equation}
\begin{array}{l}
   \vspace{1mm}
   \sum\limits_{n=1}^{+\infty} \alpha_{j}^{(n)} =
     \tilde{T}_{j-1}^{+}
     \Bigl(1 + \sum\limits_{m=1}^{+\infty} (R_{j}^{+} \tilde{R}_{j-1}^{-})^{m} \Bigr) =
     \displaystyle\frac{\tilde{T}_{j-1}^{+}} {1 - R_{j}^{+} \tilde{R}_{j-1}^{-}} =
     \displaystyle\frac{\tilde{T}_{j}^{+}} {T_{j}^{+}}, \\

   \vspace{1mm}
   \sum\limits_{n=1}^{+\infty} \beta_{j}^{(n)} =
     \tilde{T}_{j-1}^{+}
     \Bigl(1 + \sum\limits_{m=1}^{+\infty} (\tilde{R}_{j}^{+} \tilde{R}_{j-1}^{-})^{m} \Bigr)\, R_{j}^{+} =
     \displaystyle\frac{\tilde{T}_{j-1}^{+}\, R_{j}^{+}} {1 - \tilde{R}_{j}^{+} \tilde{R}_{j-1}^{-}} =
     \displaystyle\frac{\tilde{T}_{j}^{+} R_{j}^{+}} {T_{j}^{+}},
\end{array}
\label{eq.3.2.7}
\end{equation}
the amplitudes of transmission and reflection:
\begin{equation}
\begin{array}{cccc}
  A_{T} = \sum\limits_{n=1}^{+\infty} A_{T}^{(n)}, &
  A_{R} = \sum\limits_{n=1}^{+\infty} A_{R}^{(n)}, &
  \alpha_{j} = \sum\limits_{n=1}^{+\infty} \alpha_{j}^{(n)} = \displaystyle\frac{\tilde{T}_{j}^{+}} {T_{j}^{+}}, &
  \beta_{j} = \sum\limits_{n=1}^{+\infty} \beta_{j}^{(n)} = \alpha_{j} \cdot R_{j}^{+}
\end{array}
\label{eq.3.2.8}
\end{equation}
and coefficients $T$ and $R$ describing penetration of the packet from the internal region outside
and its reflection from the barrier
\begin{equation}
\begin{array}{ll}
\vspace{1mm}
  T_{MIR} \equiv \displaystyle\frac{k_{N}}{k_{M}}\; \bigl|A_{T}\bigr|^{2} =
    \bigl|A_{\rm inc}\bigr|^{2} \cdot T_{\rm bar}, &
  T_{\rm bar} = \displaystyle\frac{k_{N}}{k_{M}}\; \bigl|\tilde{T}_{N-1}^{+} \bigr|^{2}, \\
  R_{MIR} \equiv \bigl|A_{R}\bigr|^{2} = \bigl|A_{\rm inc}\bigr|^{2} \cdot R_{\rm bar}, &
  R_{\rm bar} = \bigl|\tilde{R}_{M}^{+} \bigr|^{2}.
\end{array}
\label{eq.3.2.9}
\end{equation}
Choosing $a_{\rm min}=0$, we assume full propagation of the packet through such a boundary (without any possible reflection) and we have $R_{0}^{-} = -1$ (it could be interesting to analyze results varying $R_{0}^{-}$).
We check the property:
\begin{equation}
\begin{array}{ccc}
  \displaystyle\frac{k_{N}}{k_{M}}\; |A_{T}|^{2} + |A_{R}|^{2} = 1 & \mbox{ or }&
  T_{MIR} + R_{MIR} = 1,
\end{array}
\label{eq.3.2.10}
\end{equation}
which should be the test, whether the method MIR gives us proper solution for wave function. Now if energy of the packet is located below then height of one step with number $m$, then for description of transition of this packet through such barrier
with its tunneling it shall need to use the following change:
\begin{equation}
  k_{m} \to i\,\xi_{m}.
\label{eq.3.2.11}
\end{equation}
For the barrier consisting from two rectangular steps of arbitrary heights and widths we have already obtained coincidence between amplitudes calculated by method of MIR and the corresponding amplitudes found by standard approach of quantum mechanics up to first 15 digits. Increasing of number of steps up to some thousands keeps such accuracy and fulfillment of the property (\ref{eq.3.2.10}) (see Appendix~\ref{sec.app.1} where we present shortly the standard technique of quantum mechanics applied for the potential (\ref{eq.3.2.1}) and all obtained amplitudes). This is important test which confirms reliability of the method MIR. So, we have obtained full coincidence between all amplitudes, calculated by method MIR and by standard approach of quantum mechanics, and that is way we generalize the method MIR for description of tunneling of the packet through potential, consisting from arbitrary number of rectangular barriers and wells of arbitrary sizes.
% *******************************************************************************************************************

% *******************************************************************************************************************
\section{Results
\label{sec.4}}

\begin{figure}[h]
\centerline{
\includegraphics[width=93mm]{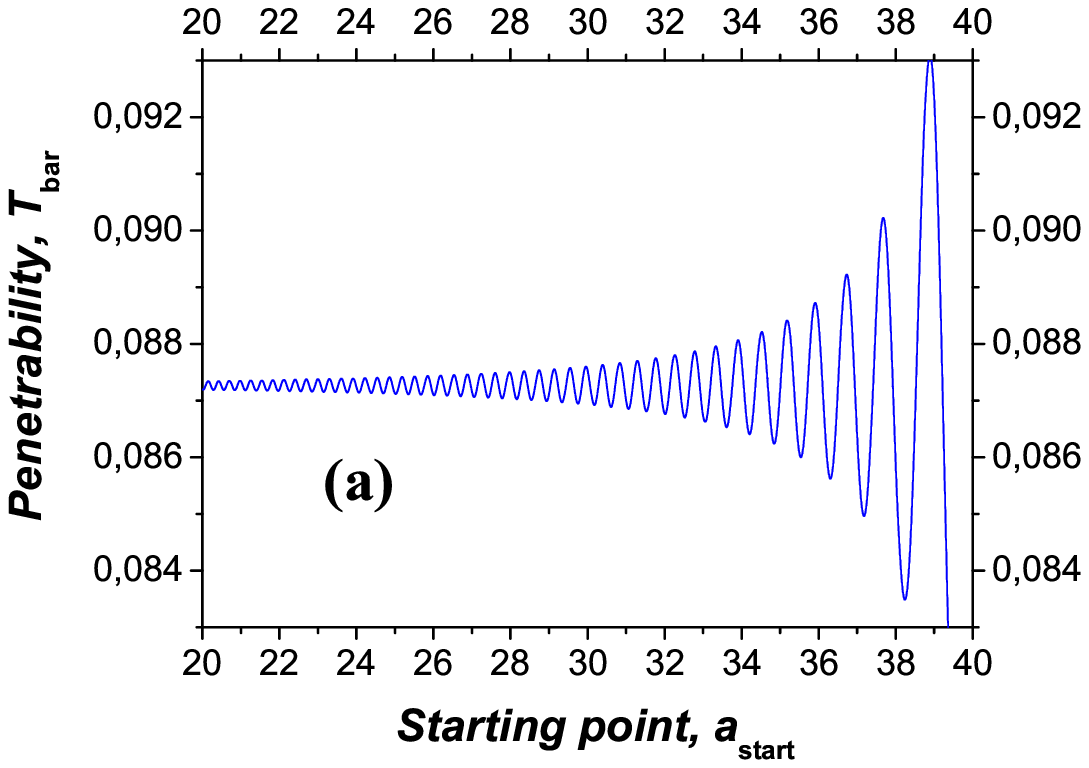}
\hspace{-7mm}\includegraphics[width=93mm]{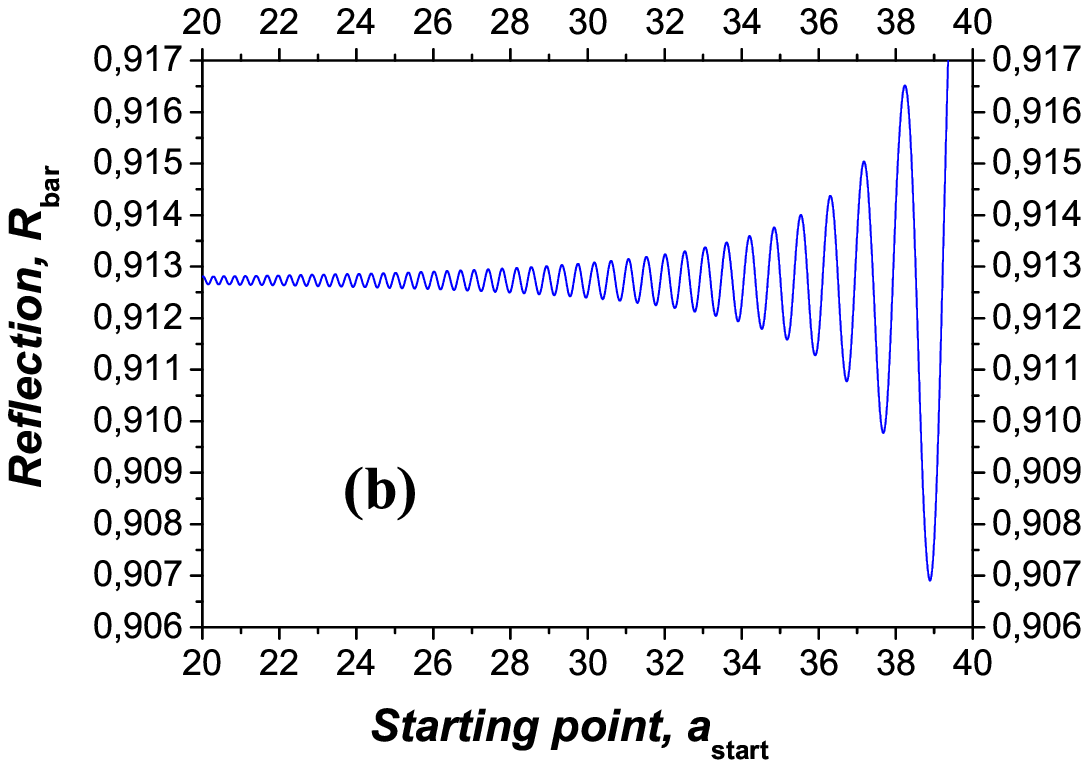}}
\vspace{-12mm}
\centerline{\includegraphics[width=93mm]{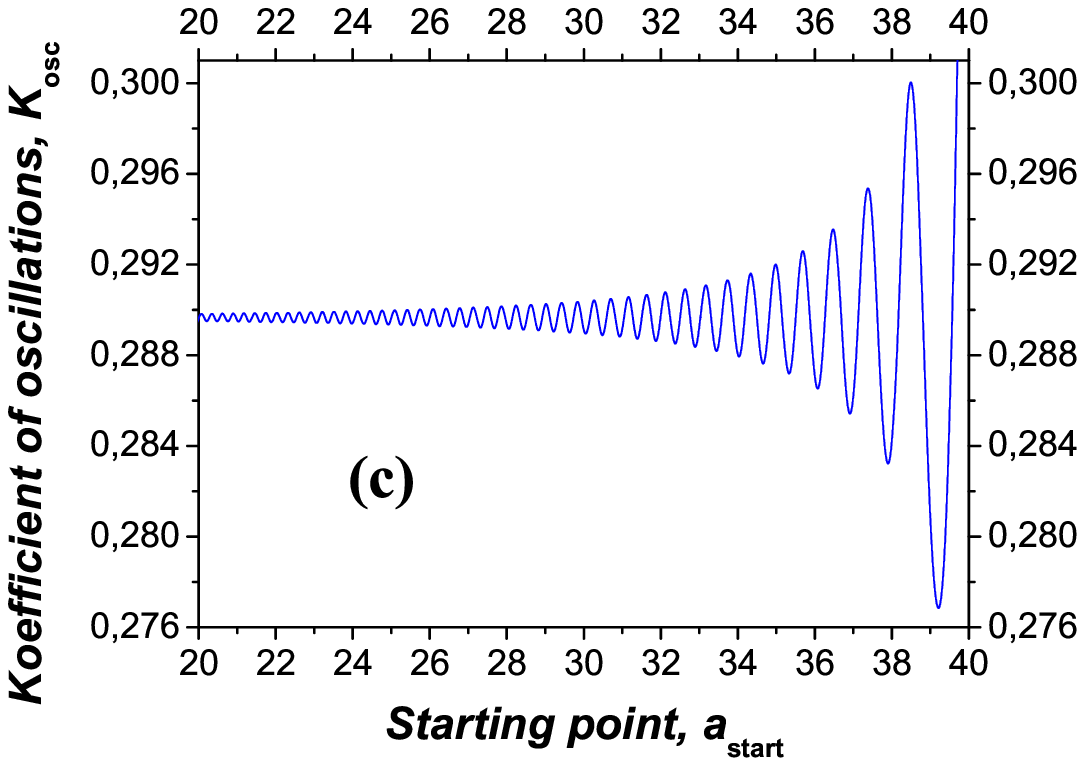}
\hspace{-7mm}\includegraphics[width=93mm]{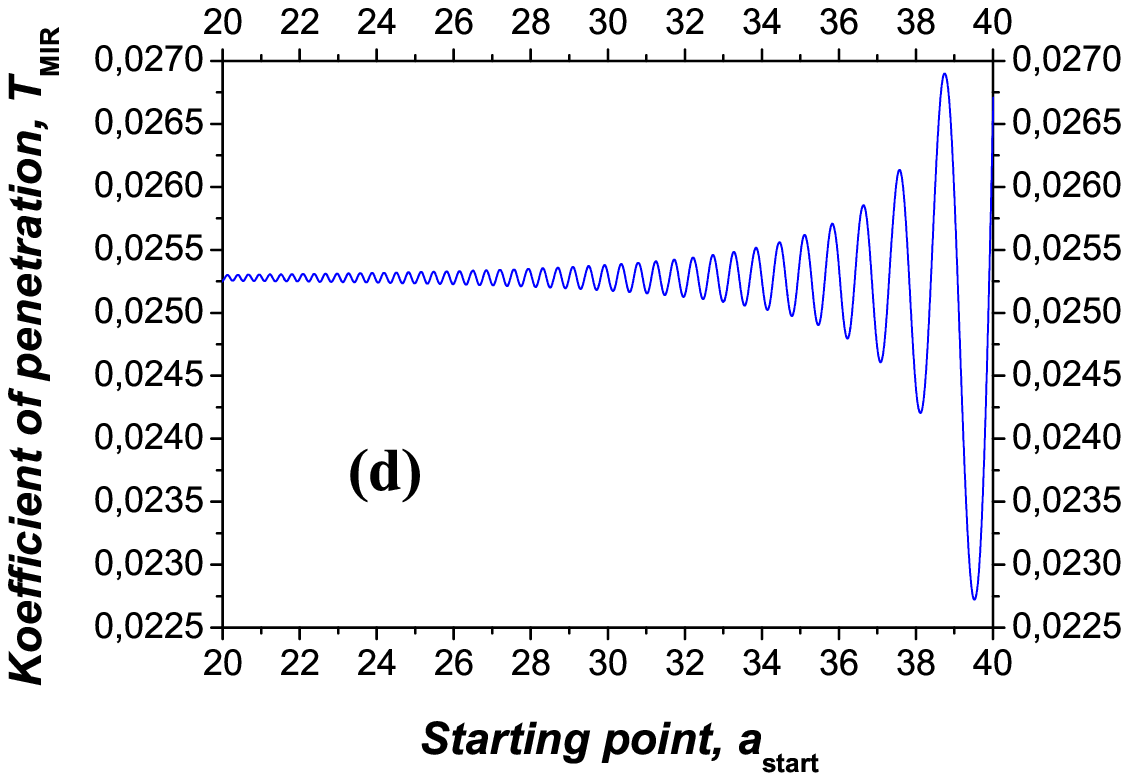}}
\vspace{-4mm}
\caption{\small
Dependencies of the coefficients of penetrability, reflection, oscillations and penetration on the position of the starting point, $a_{\rm start}$ for the energy $E=220$
($A=0.001, B=0.001$, $a_{\rm max} = 70$, total number of intervals is 2000):
(a) penetrability;
(b) reflection;
(c) coefficient of oscillation;
(d) coefficient of penetration.
For all presented values we have achieved accuracy $|T_{\rm bar}+R_{\rm bar}-1| < 1 \cdot 10^{-15}$.
\label{fig.model_Monerat.2}}
\end{figure}

We have applied the method above to analysis of behavior of the packet in its propagation relatively the barrier (\ref{eq.model.2.2}) (taking into account $a_{\rm new} \to \sqrt{12}\,a_{\rm old}$). The first interesting results which we have obtained is \emph{visible change of the penetrability on displacement of the starting point $R_{\rm min} \le r \le R_{1}$, where we putted the packet for start}. Using possibility to decrease width of intervals up to enough small limit (and choosing, for conveniens, the width of each interval to be the same), we call $R_{\rm min}$ \emph{starting point}, from where the packet begins its propagation outside. We have analyzed how much position of such a point influences on the penetrability. In Fig.~\ref{fig.model_Monerat.2} one can see that at arbitrary fixed energy of radiation $E_{\rm rad}$ the penetrability of the barrier is changed strongly in dependence on $R_{\rm min}$: it has oscillating behavior, difference between its minimums and maximums is minimal at $R_{\rm min}$ in the center of the well (i.~e. its change tends to zero in the center of the well), with increasing of $R_{\rm min}$ this difference increases, achieving to the maximum close to the turning point. On such a basis we (for the first time) establish \emph{dependence of penetrability on the starting point $R_{\rm start}$ of the packet.}
Behavior of the coefficients of reflection, oscillations and penetration turn out to be similar.

Usually, in cosmological quantum models the penetrability is determined by the barrier shape. While in non-stationary approach one can find papers where a role of the initial condition is analyzed in calculations of rates, penetrability etc. (such papers are very rare and questions about dynamics have not been studied deeply), then stationary limit does not give us a choice for work, practically. Note that such an understanding about the penetrability of the barrier is prevailing till present day not only in quantum cosmology, but also in problems of decays and fission in theoretical nuclear physics. In last topic, such a point of view is as much deep-rooted as description of dynamics of decays and fission, calculations of half-lives are fulfilled without inclusion of such initial condition. Let us give only some examples. In Ref.~\cite{Buck.1993.ADNDT} agreement between experimental data of $\alpha$-decay half-lives and ones calculated by theory is demonstrated in a wide region of nuclei from $^{106}{\rm Te}$ up to nucleus with $A_{d}=266$ and $Z_{d}=109$ (see Ref.~\cite{Denisov.2005.PHRVA,Khudenko.2009.ADNDT} for improved approaches and data of last years). In review~\cite{Sobiczewski.2007.PPNP} methodology of calculation of half-lives for spontaneous-fission is presented (see eqs.~(21)--(24) in p.~321). In prevailing two-potential approach (TPA) and semiclassical approach (WKBA) for determination of half-lives of nuclear decay by emission of proton (nuclear proton decay), which today are recognized as the most accurate and reliable (for example, see~Refs.~\cite{Gurvitz.1987.PRL,Buck.1988.PRC,Aberg.1997.PRC,Gurvitz.2004.PRA}), influence of such initial condition on results has not been taken into account and has not been analyzed yet (for details and explanations, see \cite{Maydanyuk.arXiv:0906.4739}). Such approaches forms grounds for producing tables of nuclear data (for example, see~\cite{www-nds}). The unexpected result in Fig.~\ref{fig.model_Monerat.2} leads to inevitable change of understanding about the penetrability. The first and direct conclusion is: the penetrability should be connected with the initial condition (not only in non-stationary consideration, but also in stationary one), which sets localization of start of the packet outside.
\begin{figure}[h]
\centerline{
\includegraphics[width=93mm]{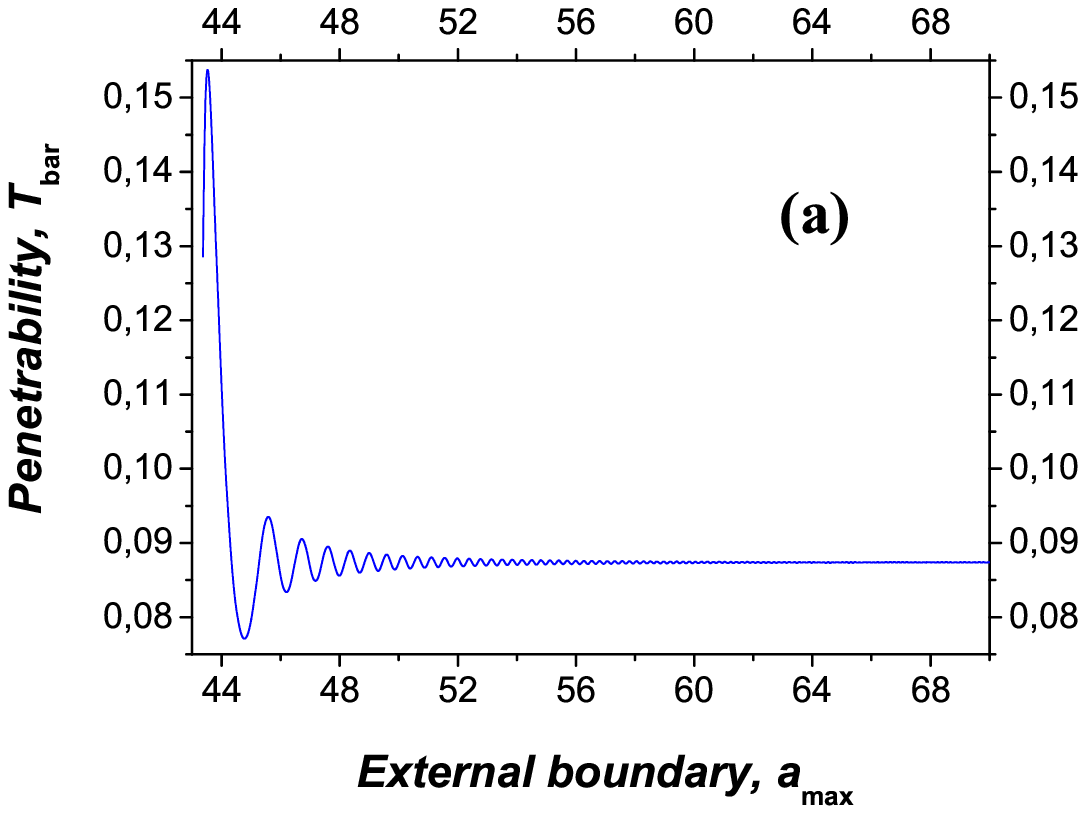}
\hspace{-7mm}\includegraphics[width=93mm]{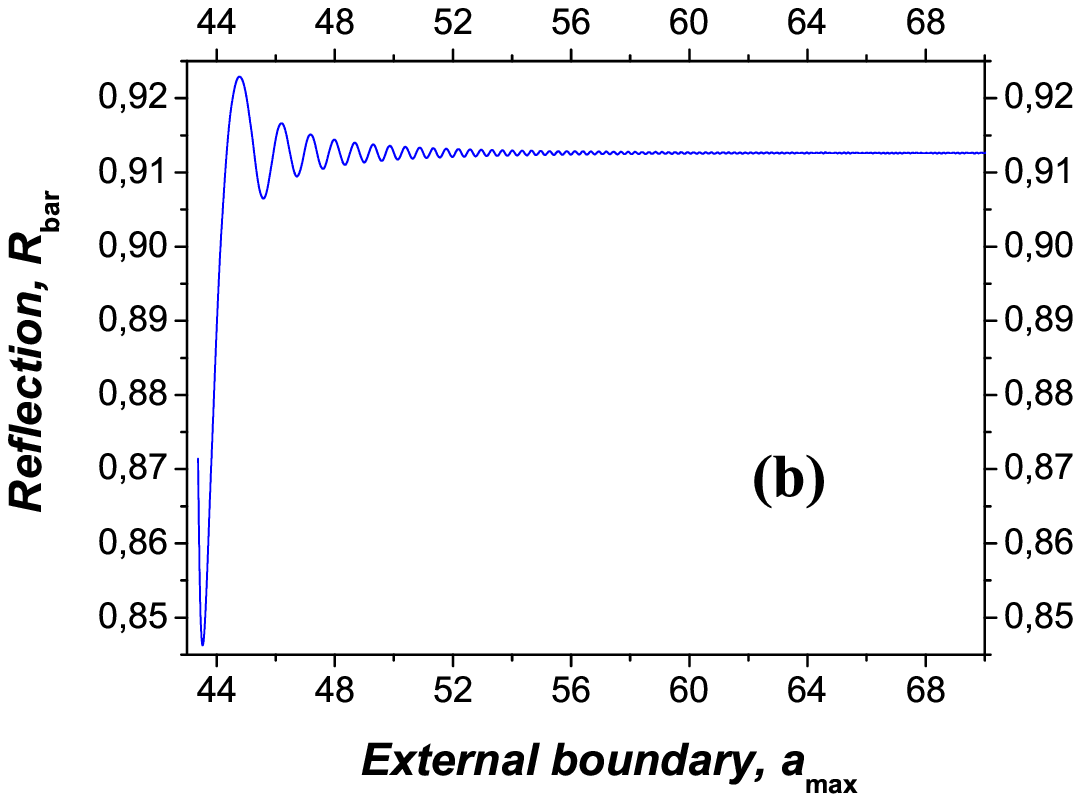}}
\vspace{-12mm}
\centerline{\includegraphics[width=93mm]{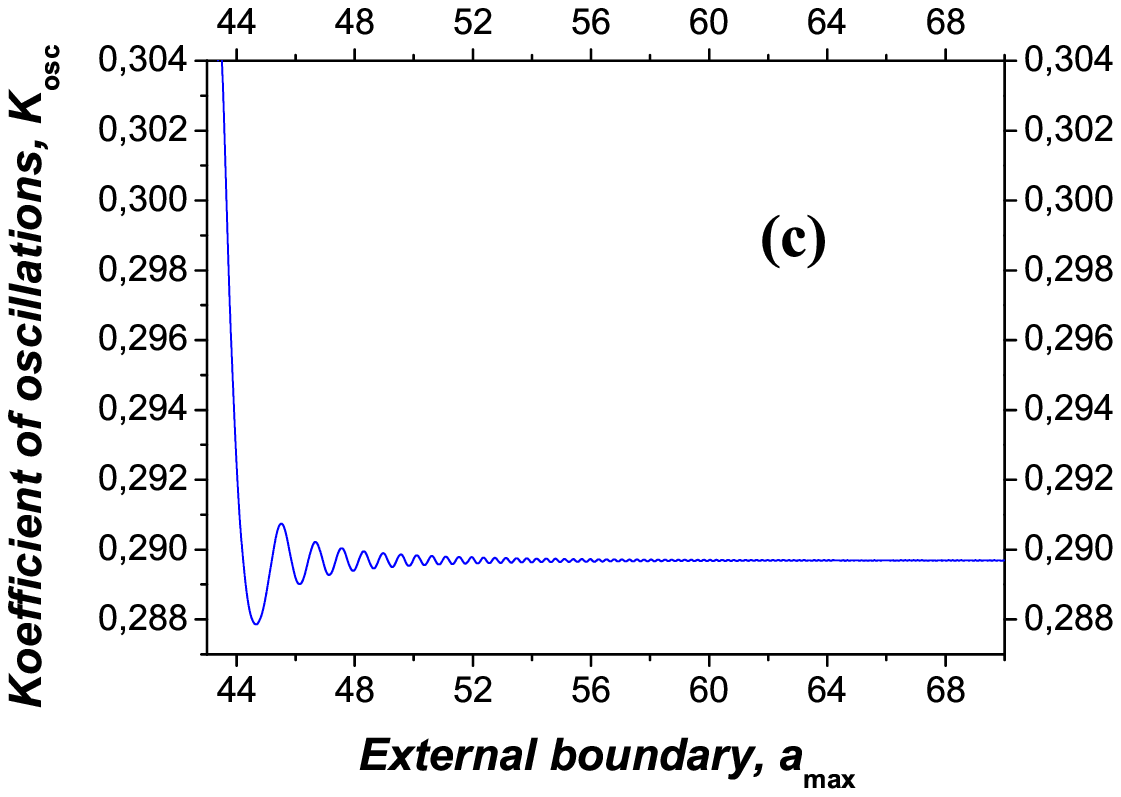}
\hspace{-7mm}\includegraphics[width=93mm]{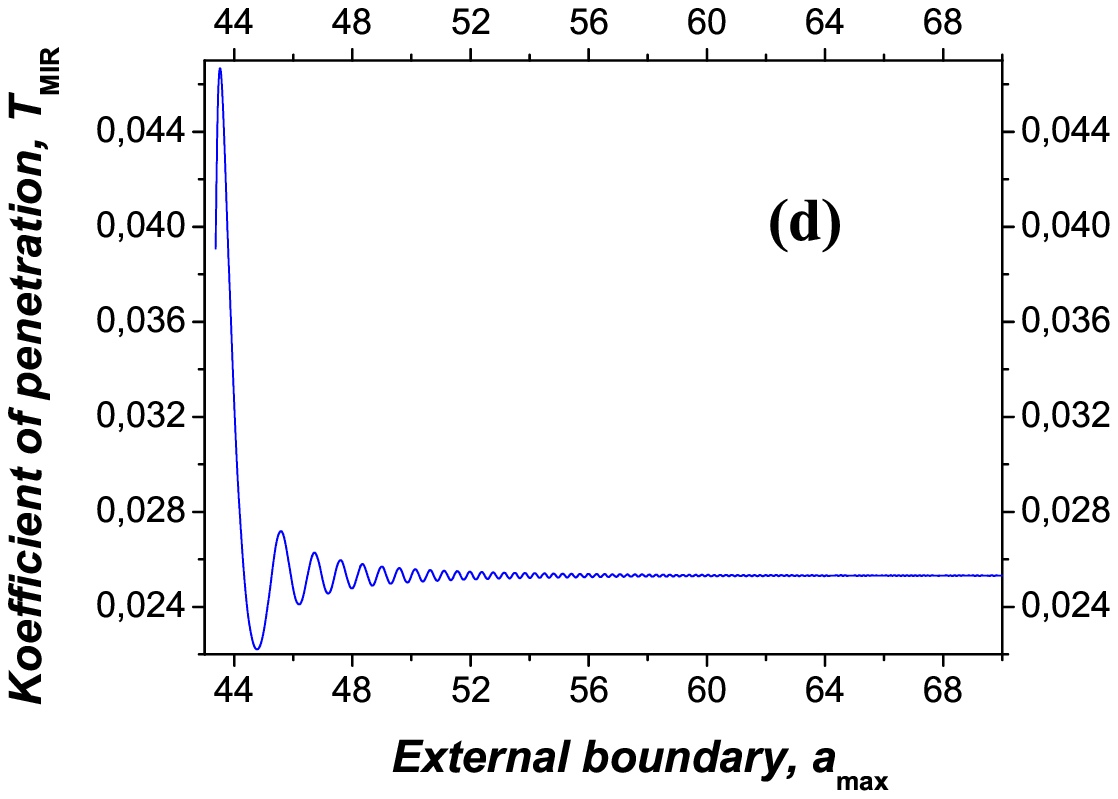}}
\vspace{-4mm}
\caption{\small
Dependencies of the coefficients of penetrability, reflection, oscillations and penetration on the position of the external region, $a_{\rm max}$ for the energy $E=223$ ($A=0.001, B=0.001$):
(a) penetrability;
(b) reflection;
(c) coefficient of oscillation;
(d) coefficient of penetration.
For all presented values we have achieved accuracy $|T_{\rm bar}+R_{\rm bar}-1| < 1 \cdot 10^{-15}$
(maximum number of intervals is 2000).
\label{fig.model_Monerat.3}}
\end{figure}
The second important conclusion is: even in the stationary consideration the penetrability of the barrier should be determined in dependence on the initial condition.

The first natural question, which could appear from analysis of such results, is how much they are reliable. In particular, would such results be destroyed if we shifted the external boundary outside (while in the semiclassical case we are restricted by two turning points only, then in the fully quantum approach the external tail of the barrier effects on the results inevitably and additionally)? Taking strong decreasing of the external tail of the barrier into account to minus infinity, one could even expect for this. Results of such calculations are presented in Fig.~\ref{fig.model_Monerat.3}, where it is shown how the penetrability is changed at increasing the external boundary $a_{\rm max}$ (for the most clearness, we have fixed the starting point: $a_{\rm start}=10$).
\begin{figure}[h]
\centerline{\includegraphics[width=93mm]{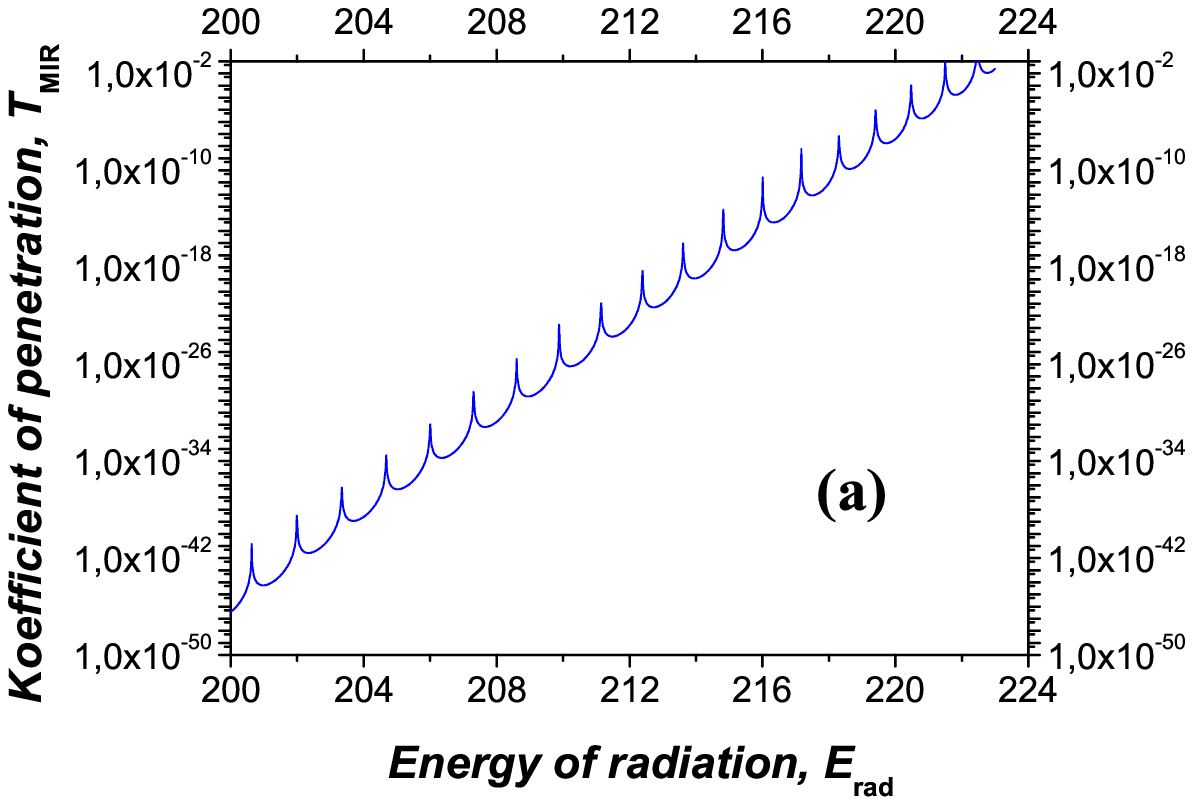}
\hspace{-4mm}\includegraphics[width=93mm]{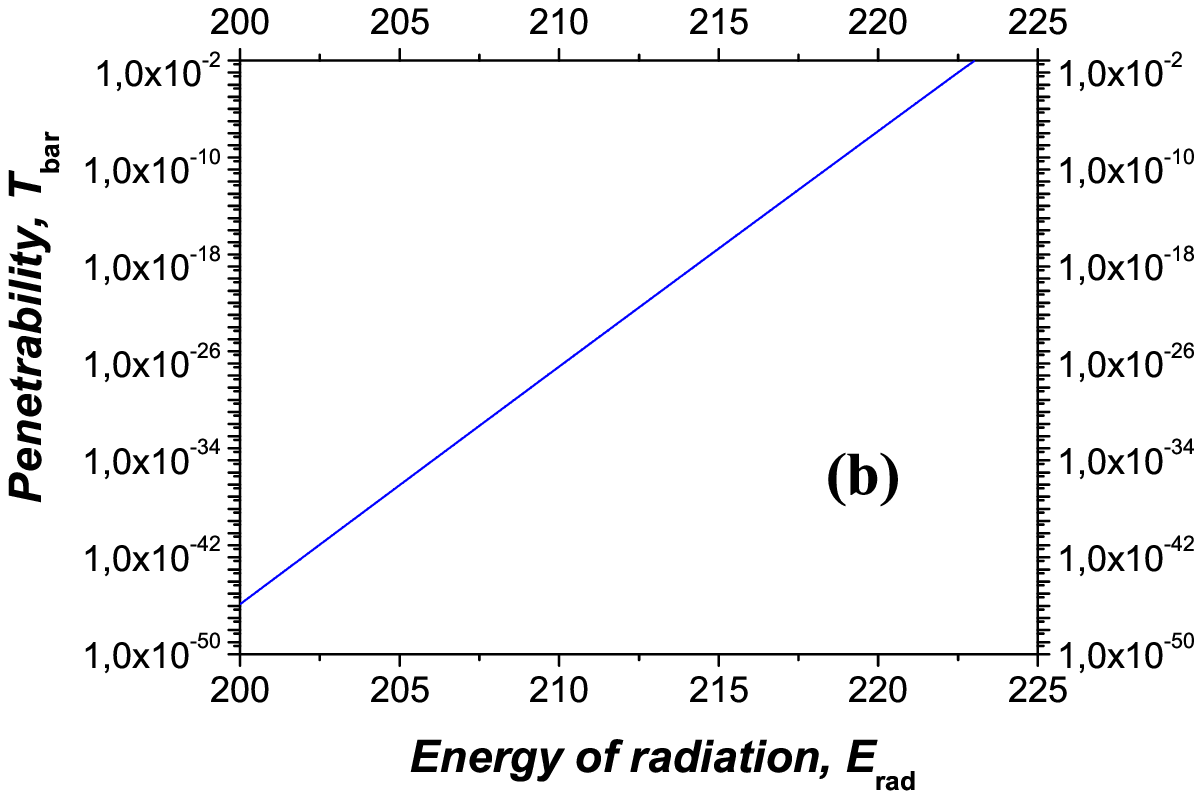}}
\vspace{-12mm}
\centerline{\includegraphics[width=93mm]{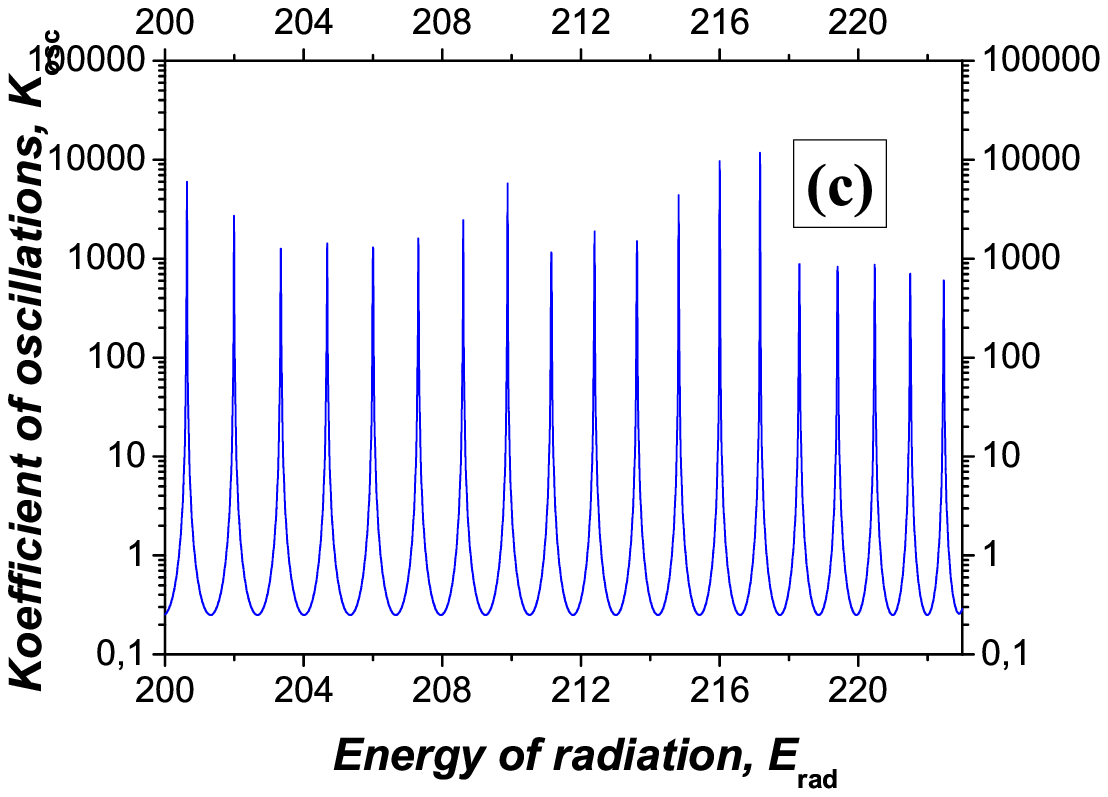}
\hspace{-4mm}\includegraphics[width=93mm]{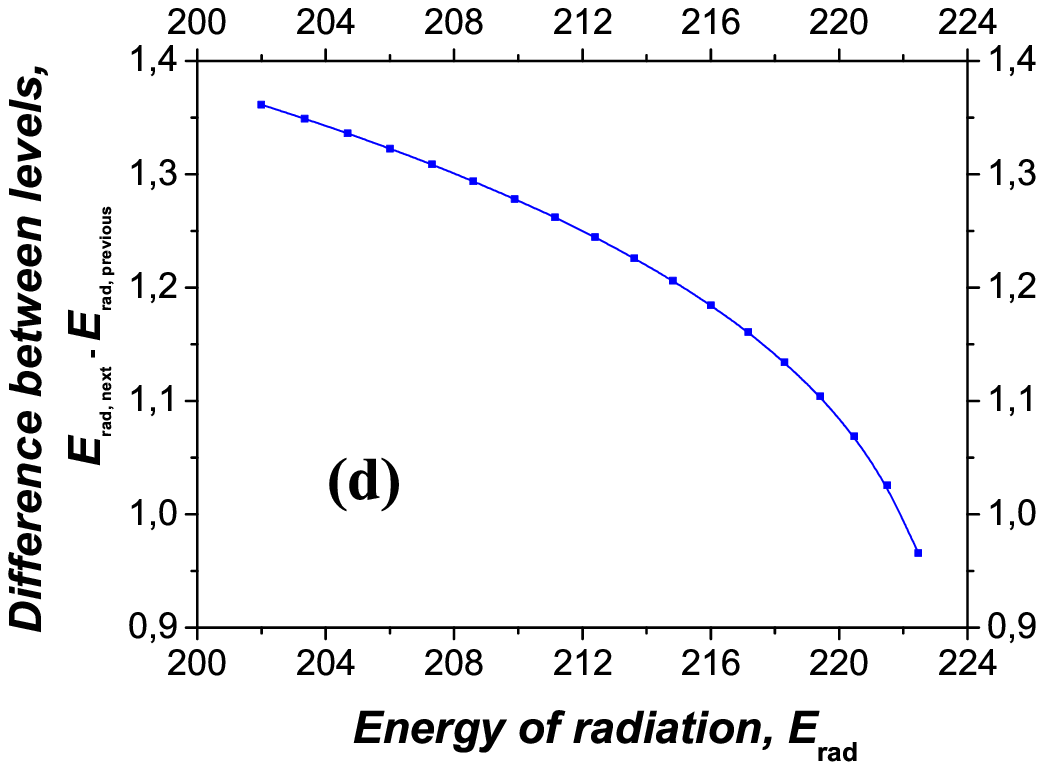}}
\vspace{-5mm}
\caption{\small
Inside the energy region $E_{\rm rad} = 200 - 223$ we have observed 19 resonant levels
(we have choose: $a_{\rm start}=10$, $a_{\rm max} = 70$, number of intervals inside the scale axis $a$ is 1000, number of intervals of energy is 100000):
(a) the coefficient of penetration in dependence of the $E_{\rm rad}$ energy;
(b) the penetrability in dependence of the $E_{\rm rad}$ energy;
(c) the coefficient of oscillation in dependence of the $E_{\rm rad}$ energy;
(d) difference between two closest resonances $E_{\rm res,\, next} - E_{\rm res,\, previous}$ in dependence on the $E_{\rm res}$ energy.
\label{fig.model_Monerat.4}}
\end{figure}
One can see that all calculations are perfectly convergent, that confirms efficiency of the techniques of the multiple internal reflections. This points to that difficult attempts and aspiration to determine the outgoing wave in the asymptotic limit properly (on which the boundary condition of tunneling is based) have no any practical sense. Therefore, we have chosen $a_{\rm max}=70$ for further calculations with the studied potential. However, one can see that inclusion of the external region in calculations can change the coefficients of penetrability and penetration up to 2 times for the chosen energy level. In Tabl.~\ref{table.model_Monerat.1} one can see how the coefficients of the penetrability and reflection are changed for different values of the starting point for the same energy. One can see that, in contrast to other known approaches, our developed method in determination of these coefficients gives the fantastic accuracy: $|T_{\rm bar}+R_{\rm bar}-1| < 1 \cdot 10^{-15}$.

The second question of not less importance is how large this effect in calculations of the penetrability: if it was small than, for example, the semiclassical approaches would have enough good approximation. From Tabl.~\ref{table.model_Monerat.1} it follows that the penetrability is changed not strongly in dependence on shift of the starting point. However, such small variations are connected with relatively small height of the barrier and depth of the well, while they would be not small at another choice of parameters (the coefficient of oscillation and penetration turn out to be changed essentially at some definite energies of radiation, see below). So, this effect is supposed to be larger at increasing height of the barrier and depth of the well, and also for near-barrier (i.~e. for energies comparable with the barrier height) and above-barrier energies of radiation.

Then, we have analyzed how these characteristics are changed in dependence on the energy of radiation. Result has turned out to be over-unexpected (see Fig.~\ref{fig.model_Monerat.4}): the coefficient of penetration has oscillating behavior where peaks are clearly shown, located about the same distances, between which smooth and stable in calculations wells are observed with minimums! By other words, in the fully quantum approach we for the first time have observed clear pictures of resonances, which could be connected with quasi-stationary states. More detailed analysis gives: at increasing energy of radiation the penetrability is changed monotonously and determines a general tendency of change of the coefficient of penetration, while the coefficient of oscillations introduces the peaks. Now a reason of presence of resonances has become clearer: oscillations of the packet inside the internal well cause them, while the possibility of the packet to penetrate through the barrier (described by the penetrability of the barrier) does not take influence on them absolutely. In Tabl.~\ref{table.model_Monerat.2} we present the resonant levels calculated by such a way inside the energy range $E_{\rm rad}$ = 200--223, and the coefficients connected with them. To such data we add a full list of the found resonant levels inside energy range $E_{\rm rad}$ = 0--200 in Tabl.~\ref{table.model_Monerat.3} (we have established 134 resonant levels). We seem for the first time to have obtained the quasi-stationary levels in the problem of decay in the fully quantum approach, separating the coefficient of penetration (having sense of width in theory of quantum scattering and decays) on the coefficients of penetrability and oscillations.

% *******************************************************************************************************************

% *******************************************************************************************************************
\section{Conclusions
\label{sec.conclusions}}

The new method for determination of probability of penetration of the packet from the internal well outside with its tunneling through one-dimensional barrier of arbitrary shape used in problems of quantum cosmology, is presented.
Note the following:
\begin{enumerate}
\item
The method is further development of approach of multiple internal reflections (see Refs.~\cite{Maydanyuk.2000.UPJ,Maydanyuk.2002.JPS,Maydanyuk.2002.PAST,Maydanyuk.2003.PhD-thesis,%
Maydanyuk.2006.FPL,Maydanyuk.arXiv:0805.4165,Maydanyuk.arXiv:0906.4739},
also Refs.~\cite{Fermor.1966.AJPIA,McVoy.1967.RMPHA,Anderson.1989.AJPIA,Esposito.2003.PRE}), where a process of tunneling of the packet through the barrier is considered consequently by steps of its propagation relatively to each boundary of the barrier. The method is fully quantum, allows to determine amplitudes of wave function, coefficients of penetrability and reflection relatively the barrier. For the first time \emph{exact} analytical solutions for amplitudes of the wave function, penetrability $T$ and reflection $R$ for the barrier, composed from arbitrary number $n$ of rectangular potential steps, are found. At limit $n \to \infty$ these solutions could be considered as exact limits for potential with interested barrier and internal well of arbitrary shape.

\item
Accuracy of the method in determination of penetrability $T_{\rm bar}$ and reflection $R_{\rm bar}$ is:
$|T_{\rm bar}+R_{\rm bar}-1| < 1 \cdot 10^{-15}$ (see Tabl.~\ref{table.model_Monerat.1}). Author has not found other methods achieving such accuracy in similar problems of quantum physics (with possible exception of some selected cases of exactly solvable barriers which could be obtained by methods of supersymmery).

\item
On the basis of the method the probability of penetration of the packet from the internal well outside with its tunneling through the barrier of arbitrary shape called \emph{coefficient of penetration}, is determined. It succeeds to separate that coefficient explicitly on the penetrability and new coefficient, which characterizes oscillating behavior of the packet inside the internal well and is called \emph{coefficient of oscillations}. That found for the first time formula seems to be fully quantum analogue of the semiclassical formula of $\Gamma$ width of decay in quasistationary state proposed by Gurvitz and K\"{a}lbermann in Ref.~\cite{Gurvitz.1987.PRL} (here, the coefficient of oscillations is fully quantum analogue for the semiclassical $F$ factor of formation and the coefficient of penetration is analogue for the semiclassical $\Gamma$ width).
\end{enumerate}

This method has been applied for study of properties of the packet, describing evolution of universe on the first stage in the closed Friedmann--Robertson--Walker model with quantization in the presence of the positive cosmological constant, radiation and component of generalize Chaplygin gas with potential chosen from \cite{Monerat.2007.PRD}. Let us formulate main results obtained for the first time:
\begin{enumerate}
\item
For the same chosen energy of radiation $E_{\rm rad}$ the penetrability of the barrier is changed visibly in dependence on the position of the starting point $R_{\rm start}$ inside the internal well, where the packet begins to propagate (see Fig.~\ref{fig.model_Monerat.2}): the penetrability has oscillating behavior, difference between its minimums and maximums is minimal at $R_{\rm start}$ in the center of the well, with increasing $R_{\rm start}$ this difference increases achieving maximum close to the turning point. The behavior of the coefficients of reflection, oscillations and penetration turns out to be similar. Coincidence (up to the first 15 digits) of the amplitudes of the wave function obtained by such a method, with corresponding amplitudes obtained in the standard approach of quantum mechanics (see App.~\ref{sec.app.1}) at different energies $E_{\rm rad}$ confirms that this result does not depend on a choice of the fully quantum method applied for calculations. Such a peculiarity is shown in the fully quantum non-stationary and stationary considerations and it is hidden after imposing the semiclassical restrictions.

\item
In non-stationary and stationary considerations the penetrability of the barrier should be connected with initial condition localizing start of the packet. Note that possible introduction of the initial condition into known stationary semiclassical models could change their results.

\item
The penetrability is changed visibly, if to take the external tail of the barrier into account. For example, for the barrier (\ref{eq.model.2.2}) with parameters $A=0.001$ and $B=0.001$ (see Fig.~\ref{fig.model_Monerat.1}) at the energy of radiation $E_{\rm rad} = 223$ the penetrability is changed up to 2 times (see Fig.~\ref{fig.model_Monerat.3}). If to increase the external boundary $a_{\rm max}$, all amplitudes and coefficients are convergent in calculations that confirms efficiency of the developed method.

\item
Dependence of the coefficient of penetration on the energy of radiation has oscillating behavior: here peaks are clearly shown, localized approximately on the same distances, between which smooth minimums (wells) stable in calculations are observed (see Fig.~\ref{fig.model_Monerat.4}). By other words, for the first time in the fully quantum approach we have obtained clear and stable picture of resonances, which indicate on presence of some quasistationary states. Here, with increasing of the energy of radiation the penetrability is changed monotonously and it describes a general tendency of behavior of the coefficient of penetration, while the coefficient of oscillations gives peaks. Now a reason of existence of resonances becomes clear --- oscillations of the packet inside the internal well give them. For example, for the barrier (\ref{eq.model.2.2}) with parameters $A=0.001$ and $B=0.001$ I establish 134 such resonant levels inside range $E_{\rm rad}$ = 0--223 (see Tabl.~\ref{table.model_Monerat.2},~\ref{table.model_Monerat.3}).

\item
The dependence of the penetrability on the starting point has clear established maximums and minimums. On such a basis one can suppose that the most probable start of the packet (describing start of expansion of the universe) is in one point of such maximums. This allows to predict some definite initial values of the scale factor, at which the universe begins to expand (such initial data is direct result of quantization of the classical cosmological model).

\item
Modulus of the wave function both in the internal, and in the external regions has clear established own maximums and minimums~\cite{Maydanyuk.2008.EPJC,Maydanyuk.2010.IJMPD}. This indicates on such values of the scale factor, at which ``appearance'' of the universe will be more or less probable. By other words, radius of the universe during its expansion is changed not continuously, but passes consequently through some definite discrete values connected with these maximums. It follows that after quantization space-time of universe on the first stage of its expansion seems to be rather discrete than continuous. According to results~\cite{Maydanyuk.2008.EPJC,Maydanyuk.2010.IJMPD}, difference between maximums and minimums with increasing of the scale factor $a$ is slowly smoothed and we obtain obvious for us continuous structure of the space-time at latter times. Discontinuity of space-time is direct result of quantization of cosmological model, which is shown the most strongly on the first stage of expansion and disappeared after imposition of the semiclassical approximations.
\end{enumerate}
% *******************************************************************************************************************

% *******************************************************************************************************************
\appendix
\section{Direct method
\label{sec.app.1}}

We shall add shortly solution for amplitudes of the wave function obtained by standard technique of quantum mechanics which could be obtained if to use only condition of continuity of the wave function and its derivative at each boundary, but on the whole region of the studied potential. At first, we find functions $f_{2}$ and $g_{2}$
(from the first boundary):
\begin{equation}
\begin{array}{cc}
  f_{2} = \displaystyle\frac{k_{2}+k}{k_{2}-k} \,e^{2ik_{2}x_{1}}, &
  g_{2} = \displaystyle\frac{2k}{k-k_{2}} \,e^{i(k+k_{2})x_{1}}.
\end{array}
\label{eq.app.3.2.1}
\end{equation}
Then, using the following recurrent relations:
\begin{equation}
\begin{array}{ccl}
  f_{j+1} & = & \displaystyle\frac
              {(k_{j+1}-k_{j})\, e^{2ik_{j}x_{j}} + f_{j}\, (k_{j+1}+k_{j}) }
              {(k_{j+1}+k_{j})\, e^{2ik_{j}x_{j}} + f_{j}\, (k_{j+1}-k_{j}) }
              \cdot e^{2ik_{j+1}x_{j}},
\end{array}
\label{eq.app.1.2}
\end{equation}
we calculate next functions $f_{3}$, $f_{4}$, $f_{5}$ \ldots $f_{n}$,
and by such a formula:
\begin{equation}
\begin{array}{ccl}
  g_{j+1} & = & g_{j} \cdot \displaystyle\frac{2k_{j}\, e^{i(k_{j+1}+k_{j})x_{j}}}
              {(k_{j+1}+k_{j})\, e^{2ik_{j}x_{j}} + f_{j}\, (k_{j+1}-k_{j})}
\end{array}
\label{eq.app.1.3}
\end{equation}
the functions $g_{3}$, $g_{4}$, $g_{5}$ \ldots $g_{n}$.
From $f_{n}$ and $g_{n}$ we find amplitudes $\alpha_{n}$, $\beta_{n}$ and amplitude of transmission $A_{T}$:
\begin{equation}
\begin{array}{cc}
  \beta_{n} = 0, &
  A_{T} = \alpha_{n} = -\displaystyle\frac{g_{n}} {f_{n}}.
\end{array}
\label{eq.app.1.4}
\end{equation}
Now using the recurrent relations:
\begin{equation}
\begin{array}{ccl}
  \alpha_{j-1} & = & \displaystyle\frac
          {\alpha_{j}\, e^{ik_{j}x_{j-1}} + \beta_{j}\, e^{-ik_{j}x_{j-1}} - g_{j-1}\, e^{-ik_{j-1}x_{j-1}}}
          {e^{ik_{j-1}x_{j-1}} + f_{j-1}\, e^{-ik_{j-1}x_{j-1}}}
\end{array}
\label{eq.app.1.5}
\end{equation}
and such a formula:
\begin{equation}
  \beta_{j} = \alpha_{j} \cdot f_{j} + g_{j},
\label{eq.app.1.6}
\end{equation}
we consistently calculate the amplitudes $\alpha_{n-1}$, $\beta_{n-1}$, $\alpha_{n-2}$, $\beta_{n-2}$ \ldots $\alpha_{2}$, $\beta_{2}$.
At finishing, we find amplitude of reflection $A_{R}$:
\begin{equation}
  A_{R} = \alpha_{2}\,e^{i(k+k_{2})x_{1}} + \beta_{2}\,e^{i(k-k_{2})x_{1}} - e^{2 ikx_{1}}.
\label{eq.app.1.7}
\end{equation}
As test we use condition:
\begin{equation}
% \begin{array}{cc}
  \displaystyle\frac{k_{n}}{k_{1}}\; |A_{T}|^{2} + |A_{R}|^{2} = 1.
%   & \mbox{ or } & T_{MIR} + R_{MIR} = 1.
% \end{array}
\label{eq.app.1.8}
\end{equation}
In order to check all amplitudes obtained previously by the MIR approach, we have used such a techniques and obtained coincidence up to the first 15 digits for all considered amplitudes. In particular, we reconstruct completely the pictures of the probability and reflection presented in Fig.~\ref{fig.model_Monerat.2} (a) and (b), Fig.~\ref{fig.model_Monerat.3} (a) and (b), Fig.~\ref{fig.model_Monerat.4}~(b), but using standard technique above. So, \emph{result on the oscillating dependence of the penetrability of the position of the starting point $R_{\rm form}$ in such figures is independent on the fully quantum method chosen for calculations}.

%-----------------------------------------------------------------------------------------------------------------------

%-----------------------------------------------------------------------------------------------------------------------
\subsection{Potential of Monerat et al.
\label{model.6}}

\begin{table}
\begin{center}
\begin{tabular}{|c|c|c|c|c|c|c|c|c|c|} \hline
  Starting & \multicolumn{4}{|c|}{Fully quantum method} \\ \cline{2-5}
  point, $a_{\rm start}$ &
  Penetrability, $T_{\rm bar}$ &
  Reflection, $R_{\rm bar}$ &
  Oscillation, $K_{\rm osc}$ &
  Penetration, $T_{MIR}$ \\ \hline
 0.07 & 0.0872668383565564 & 0.912733161643444 & 0.289664246314552 & 0.0252780829608057 \\
%0.14 & 0.0872669108203322 & 0.912733089179668 & 0.289662878661005 & 0.0252779846000706 \\
 0.28 & 0.0872671527519489 & 0.912732847248051 & 0.289665587084641 & 0.0252782910350983 \\
 0.42 & 0.0872651037703366 & 0.912734896229664 & 0.289664525101506 & 0.0252776048415682 \\
 0.63 & 0.0872643145451912 & 0.912735685454809 & 0.289664711594944 & 0.0252773925052633 \\
 0.84 & 0.0872635259989442 & 0.912736474001056 & 0.289664872338244 & 0.0252771781182692 \\
 1.05 & 0.0872627384436066 & 0.912737261556393 & 0.289665000055905 & 0.0252769611361457 \\
 1.54 & 0.0872687846742469 & 0.912731215325753 & 0.289674019798181 & 0.0252794996594910 \\
 2.03 & 0.08.72717363805440 & 0.912728263619456 & 0.289649350441997 & 0.0252782017545698 \\
 2.52 & 0.08.72573694115204 & 0.912742630588480 & 0.289663896712698 & 0.0252753096406404 \\
 3.01 & 0.08.72689349664266 & 0.912731065033573 & 0.289685966459819 & 0.0252805857676684 \\
 3.50 & 0.08.72775699699062 & 0.912722430030094 & 0.289642192688668 & 0.0252792667386222 \\
 4.06 & 0.08.72658901116154 & 0.912734109888385 & 0.289696448926345 & 0.0252806184777316 \\
 4.55 & 0.08.72834057554164 & 0.912716594244584 & 0.289644249841587 & 0.0252811365836465 \\
 5.04 & 0.08.72551358979240 & 0.912744864102076 & 0.289637485908461 & 0.0252723581940758 \\
 6.02 & 0.08.72881914256786 & 0.912711808574321 & 0.289672548079996 & 0.0252849928275708 \\
 7.00 & 0.08.72390652507793 & 0.912760934749221 & 0.289672862933907 & 0.0252707897908712 \\
 8.05 & 0.08.72525738834552 & 0.912747426116545 & 0.289611311229824 & 0.0252693323305645 \\
 9.03 & 0.08.72766044274669 & 0.912723395572533 & 0.289731953026223 & 0.0252868210542670 \\
10.02 & 0.08.72747803845243 & 0.912725219615476 & 0.289575702298831 & 0.0252726558228249 \\
11.00 & 0.0872276981949353 & 0.912772301805065 & 0.289737152195946 & 0.0252731048676081 \\
12.05 & 0.0872776209107824 & 0.912722379089218 & 0.289552884375544 & 0.0252714868761523 \\
13.03 & 0.0872097180587336 & 0.912790281941266 & 0.289669677006127 & 0.0252620108618688 \\
14.01 & 0.0872297525057242 & 0.912770247494276 & 0.289791629591448 & 0.0252784521274925 \\
15.06 & 0.0873336978425274 & 0.912666302157473 & 0.289578213304735 & 0.0252899361825347 \\
16.04 & 0.0873340452963740 & 0.912665954703626 & 0.289545256222507 & 0.0252871585222866 \\
17.02 & 0.0873513774626095 & 0.912648622537391 & 0.289581718367642 & 0.0252953619874030 \\
18.00 & 0.0873421130455704 & 0.912657886954430 & 0.289745254914410 & 0.0253069628091520 \\
19.05 & 0.0873621796972233 & 0.912637820302777 & 0.289708963481124 & 0.0253096065275343 \\
20.04 & 0.0871687991545100 & 0.912831200845490 & 0.289846866037466 & 0.0252656032511840 \\
20.11 & 0.0873358217697875 & 0.912664178230213 & 0.289824379615929 & 0.0253120503626760 \\
21.02 & 0.0873407863962863 & 0.912659213603714 & 0.289408078297668 & 0.0252771291479563 \\
22.00 & 0.0871537607549649 & 0.912846239245035 & 0.289904994878596 & 0.0252663105653185 \\
23.05 & 0.0873353606713408 & 0.912664639328659 & 0.289330571976127 & 0.0252687898567804 \\
24.03 & 0.0871264750697614 & 0.912873524930239 & 0.289506924291343 & 0.0252237178217930 \\
25.01 & 0.0870668214302267 & 0.912933178569773 & 0.289723176821301 & 0.0252252761004983 \\
26.06 & 0.0871325064377840 & 0.912867493562216 & 0.290117973838998 & 0.0252787062232434 \\
27.04 & 0.0870118636142125 & 0.912988136385788 & 0.289691943875087 & 0.0252066359105951 \\
28.02 & 0.0874450681412740 & 0.912554931858726 & 0.289084110255789 & 0.0252789797198770 \\
29.00 & 0.0871250184718151 & 0.912874981528185 & 0.290388261985181 & 0.0253000826894572 \\
30.06 & 0.0873086705738518 & 0.912691329426148 & 0.288817117691736 & 0.0252162385846372 \\
31.04 & 0.0868071036726108 & 0.913192896327389 & 0.289619296892063 & 0.0251410123308979 \\
32.02 & 0.0866924471284098 & 0.913307552871590 & 0.290071707190809 & 0.0251470261390868 \\
33.00 & 0.0866706519138812 & 0.913329348086119 & 0.289193475557903 & 0.0250645870558445 \\
34.05 & 0.0872532139618291 & 0.912746786038171 & 0.287841171185186 & 0.0251150672964445 \\
35.03 & 0.0875410677907989 & 0.912458932209201 & 0.291902880497247 & 0.0255534898499390 \\
36.01 & 0.0882330765799585 & 0.911766923420042 & 0.286694489516019 & 0.0252959368485190 \\
37.06 & 0.0854845848577083 & 0.914515415142292 & 0.287618039346707 & 0.0245869086911412 \\
38.04 & 0.0850571220982535 & 0.914942877901746 & 0.285142815727187 & 0.0242534272927471 \\
39.02 & 0.0918912908216789 & 0.908108709178321 & 0.279635849388475 & 0.0256960991603236 \\
40.01 & 0.0810849601477858 & 0.918915039852214 & 0.329388280906961 & 0.0267084356304886 \\
41.06 & 0.151504249218261  & 0.848495750781739 & 0.292960219409352 & 0.0443847180924309 \\
41.13 & 0.141989039457346  & 0.858010960542654 & 0.353172326894230 & 0.0501465994586273 \\
41.20 & 0.108251284827932  & 0.891748715172068 & 0.641964142658414 & 0.0694934432562353 \\
\hline
\end{tabular}
\end{center}
\vspace{-3mm}
\caption{\small
Dependencies of the coefficients of penetrability $T_{\rm bar}$, reflection $R_{\rm bar}$, oscillations $K_{\rm osc}$
and penetration $T_{MIR}$ on the position of the starting point $a_{\rm start}$ for the energy $E=223$
(we used $A=0.001, B=0.001$). For all presented values we have achieved accuracy
$|T_{\rm bar}+R_{\rm bar}-1| < 1 \cdot 10^{-15}$.
\label{table.model_Monerat.1}}
\end{table}
%-----------------------------------------------------------------------------------------------------------------------

%-----------------------------------------------------------------------------------------------------------------------
\begin{table}
\begin{center}
\begin{tabular}{|c|c|c|c|c|} \hline
  Energy, & \multicolumn{3}{|c|}{Fully quantum method} & $\Delta$\, Energy, \\ \cline{2-4}
  $E_{\rm rad}$ &
  Penetrability, $T_{\rm bar}$ &
  Oscillation, $K_{\rm osc}$ &
  Penetration, $T_{MIR}$ &
  $E_{\rm rad,\, next} - E_{\rm rad,\, prev}$ \\ \hline
% \vspace{2mm}
200.631586315863 & $2.40152454 \times 10^{-45}$ & $1.64916505 \times 10^{+6}$ &
 $3.96051035 \times 10^{-39}$ & - \\
201.993199931999 & $1.14426158 \times 10^{-42}$ & $1.89183111 \times 10^{+6}$ &
 $2.16474967 \times 10^{-36}$ & 1.36161361613616 \\
203.342393423934 & $5.14289866 \times 10^{-40}$ & $1.45030359 \times 10^{+6}$ &
 $7.45876441 \times 10^{-34}$ & 1.34919349193491 \\
204.678706787068 & $2.15097704 \times 10^{-37}$ & $6.62174746 \times 10^{+5}$ &
 $1.42432267 \times 10^{-31}$ & 1.33631336313363 \\
206.001450014500 & $8.44746902 \times 10^{-35}$ & $6.74278273 \times 10^{+5}$ &
 $5.69594482 \times 10^{-29}$ & 1.32274322743227 \\
207.310163101631 & $3.08154101 \times 10^{-32}$ & $4.64608634 \times 10^{+5}$ &
 $1.43171056 \times 10^{-26}$ & 1.30871308713087 \\
208.603926039260 & $1.04396547 \times 10^{-29}$ & $1.62095089 \times 10^{+6}$ &
 $1.69221676 \times 10^{-23}$ & 1.29376293762937 \\
209.882278822788 & $3.28814440 \times 10^{-27}$ & $2.49850310 \times 10^{+6}$ &
 $8.21543901 \times 10^{-21}$ & 1.27835278352783 \\
211.144301443014 & $9.50957150 \times 10^{-25}$ & $1.65795995 \times 10^{+6}$ &
 $1.57664887 \times 10^{-18}$ & 1.26202262022620 \\
212.389073890739 & $2.54399190 \times 10^{-22}$ & $3.64786127 \times 10^{+5}$ &
 $9.28012954 \times 10^{-17}$ & 1.24477244772447 \\
213.614986149861 & $6.19406186 \times 10^{-20}$ & $9.03599338 \times 10^{+5}$ &
 $5.59695021 \times 10^{-14}$ & 1.22591225912259 \\
214.821118211182 & $1.37175187 \times 10^{-17}$ & $5.19487954 \times 10^{+5}$ &
 $7.12608576 \times 10^{-12}$ & 1.20613206132061 \\
216.005400054000 & $2.75087788 \times 10^{-15}$ & $2.15778829 \times 10^{+6}$ &
 $5.93581210 \times 10^{-9}$ & 1.18428184281842 \\
217.166221662217 & $4.91371044 \times 10^{-13}$ & $4.16964758 \times 10^{+5}$ &
 $2.04884408 \times 10^{-7}$ & 1.16082160821608 \\
218.300363003630 & $7.77746865 \times 10^{-11}$ & $7.48378112 \times 10^{+5}$ &
 $5.82048731 \times 10^{-5}$ & 1.13414134141341 \\
219.404604046040 & $1.07339618 \times 10^{-8}$ & $3.58886110 \times 10^{+5}$ &
 $3.85226983 \times 10^{-3}$ & 1.10424104241042 \\
220.473424734247 & $1.25345583 \times 10^{-6}$ & $8.94224185 \times 10^{+5}$ &
 $1.12087052 \times 10^{+0}$ & 1.06882068820688 \\
221.499004990050 & $1.20673982 \times 10^{-4}$ & $3.66509631 \times 10^{+5}$ &
 $4.42281767 \times 10^{+1}$ & 1.02558025580255 \\
222.464784647846 & $8.80196772 \times 10^{-3}$ & $4.20824526 \times 10^{+4}$ &
 $3.70408389 \times 10^{+2}$ & 0.96577965779657 \\
\hline
\end{tabular}
\end{center}
\caption{\small
The resonant levels inside the energy region $E_{\rm rad}$ = 200 - 223 (we have choose: $a_{\rm start}=10$, $a_{\rm max} = 70$, number of intervals inside the scale axis $a$ is 1000, number of intervals of energy is 100000).
We also add the coefficients of penetrability $T_{\rm bar}$, oscillations $K_{\rm osc}$ and penetration $T_{MIR}$ for such levels (we used $A=0.001, B=0.001$). For all presented energy levels, $E_{\rm rad}$, we have achieved accuracy in calculations of $T_{\rm bar}$ and $R_{\rm bar}$: $|T_{\rm bar}+R_{\rm bar}-1| < 1 \cdot 10^{-15}$.
\label{table.model_Monerat.2}}
\end{table}
%-----------------------------------------------------------------------------------------------------------------------

%-----------------------------------------------------------------------------------------------------------------------

\begin{table}
%\hspace{-10mm}
%\begin{center}
%
\hspace{-15mm}
\begin{minipage}[t]{64mm}
\begin{tabular}{|c|c|c|c|} \hline
  No & Energy & Penetrability, & $\Delta\, E$ \\
     & $E_{\rm rad}$ & $T_{\rm bar}$ & \\ \hline
1  &  0.1221 & $1.696 \times 10^{-518}$ & - \\
2  &  1.5261 & $9.595 \times 10^{-512}$ & 1.526 \\
3  &  3.5352 & $3.657 \times 10^{-504}$ & 2.009 \\
4  &  5.5377 & $3.908 \times 10^{-497}$ & 2.002 \\
5  &  7.5357 & $1.938 \times 10^{-490}$ & 1.998 \\
6  &  9.5293 & $5.550 \times 10^{-484}$ & 1.993 \\
7  & 11.5184 & $1.016 \times 10^{-477}$ & 1.989 \\
8  & 13.5031 & $1.290 \times 10^{-471}$ & 1.984 \\
9  & 15.4856 & $1.222 \times 10^{-465}$ & 1.982 \\
10 & 17.4614 & $8.601 \times 10^{-460}$ & 1.975 \\
11 & 19.4350 & $4.822 \times 10^{-454}$ & 1.973 \\
12 & 21.4042 & $2.174 \times 10^{-448}$ & 1.969 \\
13 & 23.3689 & $7.949 \times 10^{-443}$ & 1.964 \\
14 & 25.3314 & $2.462 \times 10^{-437}$ & 1.962 \\
15 & 27.2895 & $6.413 \times 10^{-432}$ & 1.958 \\
16 & 29.2431 & $1.414 \times 10^{-426}$ & 1.953 \\
17 & 31.1945 & $2.740 \times 10^{-421}$ & 1.951 \\
18 & 33.1414 & $4.579 \times 10^{-416}$ & 1.946 \\
19 & 35.0840 & $6.686 \times 10^{-411}$ & 1.942 \\
20 & 37.0243 & $8.752 \times 10^{-406}$ & 1.940 \\
21 & 38.9579 & $9.932 \times 10^{-401}$ & 1.933 \\
22 & 40.8893 & $1.022 \times 10^{-395}$ & 1.931 \\
23 & 42.8185 & $9.533 \times 10^{-391}$ & 1.929 \\
24 & 44.7410 & $7.845 \times 10^{-386}$ & 1.922 \\
25 & 46.6614 & $5.945 \times 10^{-381}$ & 1.920 \\
26 & 48.5772 & $4.057 \times 10^{-376}$ & 1.915 \\
27 & 50.4909 & $2.551 \times 10^{-371}$ & 1.913 \\
28 & 52.3979 & $1.450 \times 10^{-366}$ & 1.906 \\
29 & 54.3027 & $7.581 \times 10^{-362}$ & 1.904 \\
30 & 56.2030 & $3.658 \times 10^{-357}$ & 1.900 \\
31 & 58.0989 & $1.617 \times 10^{-352}$ & 1.895 \\
32 & 59.9926 & $6.662 \times 10^{-348}$ & 1.893 \\
33 & 61.8796 & $2.518 \times 10^{-343}$ & 1.887 \\
34 & 63.7644 & $8.852 \times 10^{-339}$ & 1.884 \\
35 & 65.6448 & $2.902 \times 10^{-334}$ & 1.880 \\
36 & 67.5207 & $8.828 \times 10^{-330}$ & 1.875 \\
37 & 69.3944 & $2.522 \times 10^{-325}$ & 1.873 \\
38 & 71.2614 & $6.679 \times 10^{-321}$ & 1.867 \\
39 & 73.1240 & $1.639 \times 10^{-316}$ & 1.862 \\
40 & 74.9844 & $3.831 \times 10^{-312}$ & 1.860 \\
\hline
\end{tabular}
\end{minipage}
%
%\hspace{2mm}
\begin{minipage}[t]{66mm}
\begin{tabular}{|c|c|c|c|} \hline
  No & Energy & Penetrability, & $\Delta\,E$ \\
      & $E_{\rm rad}$ & $T_{\rm bar}$ & \\ \hline
41 &  76.8403 & $8.368 \times 10^{-308}$ & 1.855 \\
42 &  78.6918 & $1.709 \times 10^{-303}$ & 1.851 \\
43 &  80.5389 & $3.299 \times 10^{-299}$ & 1.847 \\
44 &  82.3793 & $5.871 \times 10^{-295}$ & 1.840 \\
45 &  84.2175 & $1.001 \times 10^{-290}$ & 1.838 \\
46 &  86.0534 & $1.623 \times 10^{-286}$ & 1.835 \\
47 &  87.8827 & $2.452 \times 10^{-282}$ & 1.829 \\
48 &  89.7076 & $3.511 \times 10^{-278}$ & 1.824 \\
49 &  91.5280 & $4.721 \times 10^{-274}$ & 1.820 \\
50 &  93.3440 & $6.050 \times 10^{-270}$ & 1.815 \\
51 &  95.1555 & $7.296 \times 10^{-266}$ & 1.811 \\
52 &  96.9626 & $8.368 \times 10^{-262}$ & 1.807 \\
53 &  98.7653 & $9.104 \times 10^{-258}$ & 1.802 \\
54 & 100.5613 & $9.263 \times 10^{-254}$ & 1.795 \\
55 & 102.3551 & $9.112 \times 10^{-250}$ & 1.793 \\
56 & 104.1444 & $8.460 \times 10^{-246}$ & 1.789 \\
57 & 105.9271 & $7.440 \times 10^{-242}$ & 1.782 \\
58 & 107.7075 & $6.267 \times 10^{-238}$ & 1.780 \\
59 & 109.4813 & $4.987 \times 10^{-234}$ & 1.773 \\
60 & 111.2507 & $3.784 \times 10^{-230}$ & 1.769 \\
61 & 113.0156 & $2.730 \times 10^{-226}$ & 1.764 \\
62 & 114.7739 & $1.870 \times 10^{-222}$ & 1.758 \\
63 & 116.5277 & $1.214 \times 10^{-218}$ & 1.753 \\
64 & 118.2771 & $7.598 \times 10^{-215}$ & 1.749 \\
65 & 120.0220 & $4.512 \times 10^{-211}$ & 1.744 \\
66 & 121.7625 & $2.578 \times 10^{-207}$ & 1.740 \\
67 & 123.4963 & $1.389 \times 10^{-203}$ & 1.733 \\
68 & 125.2257 & $7.181 \times 10^{-200}$ & 1.729 \\
69 & 126.9485 & $3.520 \times 10^{-196}$ & 1.722 \\
70 & 128.6668 & $1.649 \times 10^{-192}$ & 1.718 \\
71 & 130.3806 & $7.449 \times 10^{-189}$ & 1.713 \\
72 & 132.0878 & $3.173 \times 10^{-185}$ & 1.707 \\
73 & 133.7906 & $1.306 \times 10^{-181}$ & 1.702 \\
74 & 135.4867 & $5.073 \times 10^{-178}$ & 1.696 \\
75 & 137.1783 & $1.906 \times 10^{-174}$ & 1.691 \\
76 & 138.8633 & $6.765 \times 10^{-171}$ & 1.684 \\
77 & 140.5439 & $2.322 \times 10^{-167}$ & 1.680 \\
78 & 142.2178 & $7.550 \times 10^{-164}$ & 1.673 \\
79 & 143.8850 & $2.345 \times 10^{-160}$ & 1.667 \\
80 & 145.5478 & $6.996 \times 10^{-157}$ & 1.662 \\
\hline
\end{tabular}
\end{minipage}
%
%\hspace{8mm}
\begin{minipage}[t]{60mm}
\begin{tabular}{|c|c|c|c|} \hline
  No. & Energy & Penetrability, & $\Delta\,E$ \\
      & $E_{\rm rad}$ & $T_{\rm bar}$ & \\ \hline
 81 & 147.2040 & $1.990 \times 10^{-153}$ & 1.656 \\
 82 & 148.8534 & $5.400 \times 10^{-150}$ & 1.649 \\
 83 & 150.4985 & $1.409 \times 10^{-146}$ & 1.645 \\
 84 & 152.1368 & $3.518 \times 10^{-143}$ & 1.638 \\
 85 & 153.7686 & $8.353 \times 10^{-140}$ & 1.631 \\
 86 & 155.3936 & $1.900 \times 10^{-136}$ & 1.625 \\
 87 & 157.0120 & $4.109 \times 10^{-133}$ & 1.618 \\
 88 & 158.6260 & $8.619 \times 10^{-130}$ & 1.613 \\
 89 & 160.2310 & $1.700 \times 10^{-126}$ & 1.605 \\
 90 & 161.8295 & $3.224 \times 10^{-123}$ & 1.598 \\
 91 & 163.4234 & $5.873 \times 10^{-120}$ & 1.593 \\
 92 & 165.0085 & $1.018 \times 10^{-116}$ & 1.585 \\
 93 & 166.5870 & $1.679 \times 10^{-113}$ & 1.578 \\
 94 & 168.1587 & $2.666 \times 10^{-110}$ & 1.571 \\
 95 & 169.7216 & $3.987 \times 10^{-107}$ & 1.562 \\
 96 & 171.2801 & $5.801 \times 10^{-104}$ & 1.558 \\
 97 & 172.8274 & $7.876 \times 10^{-101}$ & 1.547 \\
 98 & 174.3704 & $1.040 \times 10^{-97}$ & 1.542 \\
 99 & 175.9044 & $1.297 \times 10^{-94}$ & 1.534 \\
100 & 177.4318 & $1.557 \times 10^{-91}$ & 1.527 \\
101 & 178.9480 & $1.750 \times 10^{-88}$ & 1.516 \\
102 & 180.4599 & $1.911 \times 10^{-85}$ & 1.511 \\
103 & 181.9606 & $1.957 \times 10^{-82}$ & 1.500 \\
104 & 183.4547 & $1.925 \times 10^{-79}$ & 1.494 \\
105 & 184.9377 & $1.782 \times 10^{-76}$ & 1.482 \\
106 & 186.4140 & $1.581 \times 10^{-73}$ & 1.476 \\
107 & 187.8814 & $1.337 \times 10^{-70}$ & 1.467 \\
108 & 189.3378 & $1.060 \times 10^{-67}$ & 1.456 \\
109 & 190.7874 & $8.123 \times 10^{-65}$ & 1.449 \\
110 & 192.2238 & $5.763 \times 10^{-62}$ & 1.436 \\
111 & 193.6535 & $3.957 \times 10^{-59}$ & 1.429 \\
112 & 195.0699 & $2.518 \times 10^{-56}$ & 1.416 \\
113 & 196.4773 & $1.534 \times 10^{-53}$ & 1.407 \\
114 & 197.8737 & $8.777 \times 10^{-51}$ & 1.396 \\
115 & 199.2590 & $4.742 \times 10^{-48}$ & 1.385 \\
116 & 200.6332 & $2.420 \times 10^{-45}$ & 1.374 \\
117 & 201.9963 & $1.160 \times 10^{-42}$ & 1.363 \\
 & & & \\
 & & & \\
 & & & \\
\hline
\end{tabular}
\end{minipage}
%\end{center}
\caption{\small
The resonant levels inside the energy region $E_{\rm rad}$ = 0 - 200 ($\Delta\, E = E_{\rm rad,\, next} - E_{\rm rad,\, prev}$, we have choose: $a_{\rm start}=10$, $a_{\rm max} = 70$, number of intervals inside the scale axis $a$ is 1000, number of intervals of energy is 100000).
We also add the coefficients of penetrability $T_{\rm bar}$, oscillations $K_{\rm osc}$ and penetration $T_{MIR}$ for such levels (we used $A=0.001, B=0.001$). For all presented energy levels, $E_{\rm rad}$, we have achieved accuracy in calculations of $T_{\rm bar}$ and $R_{\rm bar}$: $|T_{\rm bar}+R_{\rm bar}-1| < 1 \cdot 10^{-15}$.
\label{table.model_Monerat.3}}
\end{table}

% *******************************************************************************************************************

% *******************************************************************************************************************
% \newpage

\end{document}